\documentclass[11pt,preprint]{aastex}

\slugcomment{Accepted for Publication in \emph{PASP}}
\shorttitle{Accuracy of 21 cm Line Profiles with the GBT}
\shortauthors{Robishaw \& Heiles}

\begin{document}


\title{ON MEASURING ACCURATE 21 CM LINE PROFILES WITH THE ROBERT C.\ BYRD
  GREEN BANK TELESCOPE}

\author{Timothy Robishaw\footnote{Currently Postdoctoral Fellow of Radio
    Polarimetry at the Sydney Institute for Astronomy, School of Physics
    A29, The University of Sydney, NSW, 2006, Australia.}\ \ \& Carl Heiles}

\affil{Astronomy Department, University of California, Berkeley, CA
  94720-3411; robishaw@physics.usyd.edu.au, heiles@astro.berkeley.edu}


\begin{abstract}

  We use observational data to show that 21 cm line profiles measured with
  the Green Bank Telescope (GBT) are subject to significant inaccuracy.
  These include $\sim$10\% errors in the calibrated gain and significant
  contribution from distant sidelobes. In addition, there are $\sim$60\%
  variations between the GBT and Leiden/Argentine/Bonn 21 cm line profile
  intensities, which probably occur because of the high main-beam
  efficiency of the GBT. Stokes $V$ profiles from the GBT contain
  inaccuracies that are related to the distant sidelobes.

  We illustrate these problems, define physically motivated components for
  the sidelobes, and provide numerical results showing the inaccuracies.
  We provide a correction scheme for Stokes $I$ 21 cm line profiles that is
  fairly successful and provide some rule-of-thumb comments concerning the
  accuracy of Stokes $V$ profiles.

\end{abstract}




\section{INTRODUCTION}
\label{introduction}

The Robert C.\ Byrd Green Bank Telescope (GBT) has no aperture blockage from
mechanical structures such as feed legs. In conventional telescopes,
scattering from such structures produces distant sidelobes far from the
main beam, which are particularly important in contaminating 21 cm line
profiles with ``stray radiation''. For example, when observing a weak
high-latitude position with a sidelobe lying on the Galactic plane, the
profile can be heavily contaminated. The nature of the contamination
depends both on hour angle, because the distant sidelobes might hit the
ground instead of the sky, and on the observing epoch, because of Doppler
shifts from the Earth's orbital motion. A great advance in 21 cm line
survey data occurred when \citet{kalberlabhabmp05} corrected both northern
and southern sky 21 cm line survey data for stray radiation; the resulting
``LAB'' (Leiden/Argentine/Bonn) survey is largely free of stray radiation.

With its absence of aperture blockage, one would expect the GBT to have
very low stray radiation. However, this is not the case, as first shown
by \cite{lockmanc05}. We have used
the GBT to measure Zeeman splitting of the 21 cm line, which requires
long integrations. For four circumpolar positions we have nearly full
24-hour coverage. We observe systematic time variations during the
observing periods, both in Stokes $I$ and Stokes $V$. These are reminiscent
of stray radiation. The present paper explores these effects for both
Stokes $I$ and $V$ observations, and for Stokes $I$ presents an approximate
correction scheme to remove these contaminants.

We begin in \S \ref{basicprops} by describing the basic properties of the
GBT primary beam in both Stokes $I$ and $V$. However, the telescope's
response is not limited to just the primary beam: \S \ref{empiricali} shows
the contribution of stray radiation to the time variability of Stokes $I$
and $V$ 21 cm line profiles at the North Celestial Pole (NCP). \S
\ref{sidelobeobsi} presents observations of the Sun, far from the main
beam, which show distant sidelobes related to spillover from the secondary
reflector. This section defines three sidelobe components motivated by
structural and physical considerations: the spillover, the spot, and the
screen components. \S \ref{diffraction} uses observations of Cas A to
define the fourth (``nearin'') component, which is related to the
diffraction rings of the primary beam. \S \ref {lsfits} outlines the
least-squares procedure that we used to solve for the amplitude
coefficients of the four components and \S \ref{lsresults} presents the
results of the least-squares analysis for full hour-angle coverage of four
positions; these results include not only the amplitude coefficients, but
also the errors in calibrated system gains and the true profile of the
observed position. In particular, \S \ref{fitresults} shows the
contributions of the four components and how they change with time for the
four sources, and shows how well their sum---which is the total predicted
spillover profile---predicts the actual time behavior.

The first sections, summarized above, focus on Stokes $I$. \S
\ref{stokesv} treats Stokes $V$ 21 cm line profiles. This is necessarily
less detailed than the Stokes $I$ discussion, because the sidelobe
structures for Stokes $V$ are much more complicated, as shown by our
observations of the Sun and Cas A. We find that the primary contributor to
stray radiation in Stokes $V$ is the screen component, which lies close to
the main beam and has very complicated structure. Without considerable
effort, which would involve high dynamic-range observations of this
component in Stokes $V$, it is impossible to correct observed profiles.
Rather, we discuss the Stokes $V$ contamination as a fraction of the
screen's Stokes $I$ predicted profile and propose a rule of thumb for the
reliability of Stokes $V$ 21 cm line profiles measured with the GBT.

\section{BASIC PROPERTIES OF THE GBT PRIMARY BEAM AT 1420 MHZ}
\label{basicprops}

For all observations in this paper, we used the GBT and the Spectral
Processor backend \citep{fisher91}.\footnote{The National Radio Astronomy
  Observatory is a facility of the National Science Foundation operated
  under cooperative agreement by Associated Universities, Inc.}  All data
were analyzed using algorithms written by the authors in the Interactive
Data Language (IDL).

We used 3C 286 as a polarization calibrator on a number of different
occasions from 2003 January to 2006 July, observing the eight-legged
star-shaped pattern of \citet{heilespnlbghletal01}.\footnote{This pattern
  has come to be known as a ``spider scan'' for obvious reasons, and is now
  a supported observing mode at the GBT.} In the range 1--2 GHz, we
observed at four frequencies simultaneously: \{1160, 1420, 1666, 1790\}
MHz.  These observations provided the beam parameters.\footnote{For a
  primer on polarized beam properties of single-dish telescopes, see
  \citet{heilespnlbghletal01}.} \citet{heilesrtr03} reported on the results
of these observations; however, their analysis was flawed because of a
computer program error by one of us (CH), so we present updated values
here. We find: \begin{enumerate}

\item At 1420 MHz, the primary beam full width at half maximum (FWHM) is
  $9.088 \pm 0.009$ arcmin. The beamwidth is proportional to wavelength to
  within $1.7\%$ (rms) among the four frequencies.
  
\item The fractional main beam ellipticity, defined as
  \begin{equation} \label{fracellipse}
    fractional \ beam \ ellipticity =
    \frac{ {\rm FWHM}_{\rm max}-{\rm FWHM}_{\rm min} }
    { {\rm FWHM}_{\rm max}+{\rm FWHM}_{\rm min} } \ ,
  \end{equation}
  is less than $1\%$ and mostly less than $0.5\%$. 

\item The beam squint in Stokes $V$ changes systematically with zenith
  angle, ZA, and it also changes with frequency. Figure \ref{getsquint}
  exhibits this behavior for two frequencies of interest, the 21 cm line
  and the 18 cm OH lines. We define beam squint as the difference in
  central position of the two circularly polarized beams.

  \begin{figure}[!h]
    \begin{center}
      \includegraphics[width=4in]{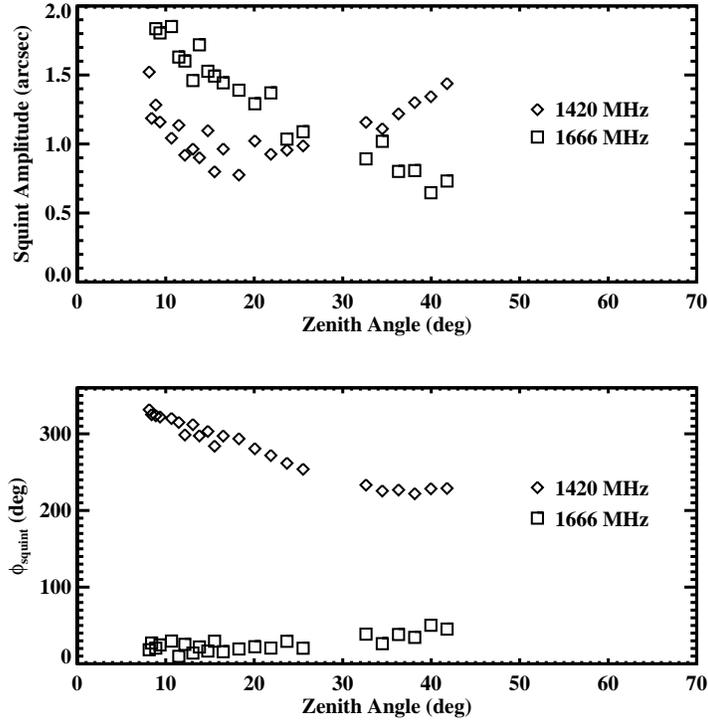}
    \end{center}
    \caption { Stokes $V$ Beam Squint versus zenith angle for 1420 and 1666
      MHz.\label{getsquint} }
  \end{figure}
  
\item The beam squash in Stokes $V$ is about 1$\arcsec$ and changes little,
  if at all, with ZA. It changes little among our three lowest frequencies,
  but triples for the highest frequency (1790 MHz). Figure \ref{getsquash}
  exhibits the squash behavior for two frequencies of interest, the 21 cm
  line and the 18 cm OH lines. We define beam squash as the difference
  between the FWHM in the orthogonal circular polarizations. Beam squash
  samples the second angular derivative of the sky brightness, and a FWHM
  difference of order 1$\arcsec$ produces essentially no detectable effect
  unless the sky brightness changes extremely rapidly with position.
  
  \begin{figure}[!h]
    \begin{center}
      \includegraphics[width=4in]{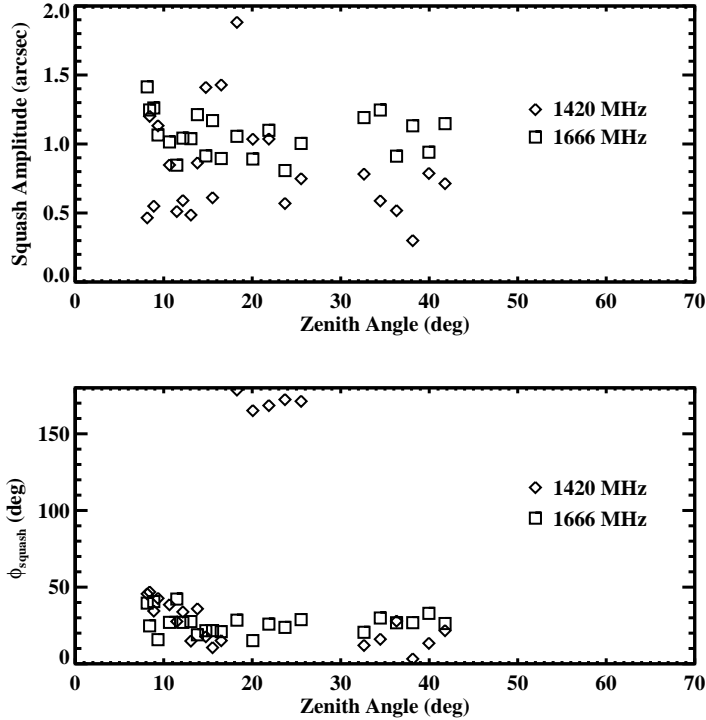}
    \end{center}
    \caption { Stokes $V$ Beam Squash versus zenith angle for 1420 and 1666
      MHz.\label{getsquash} }
  \end{figure}

\end{enumerate}

\section{EMPIRICAL EVIDENCE FOR STRAY RADIATION:
  TIME-DEPENDENT STOKES $I$ AND $V$ 21 CM LINE PROFILES}
\label{empiricali}

\begin{figure}[!p]
  \begin{center}
    \includegraphics[width=5in]{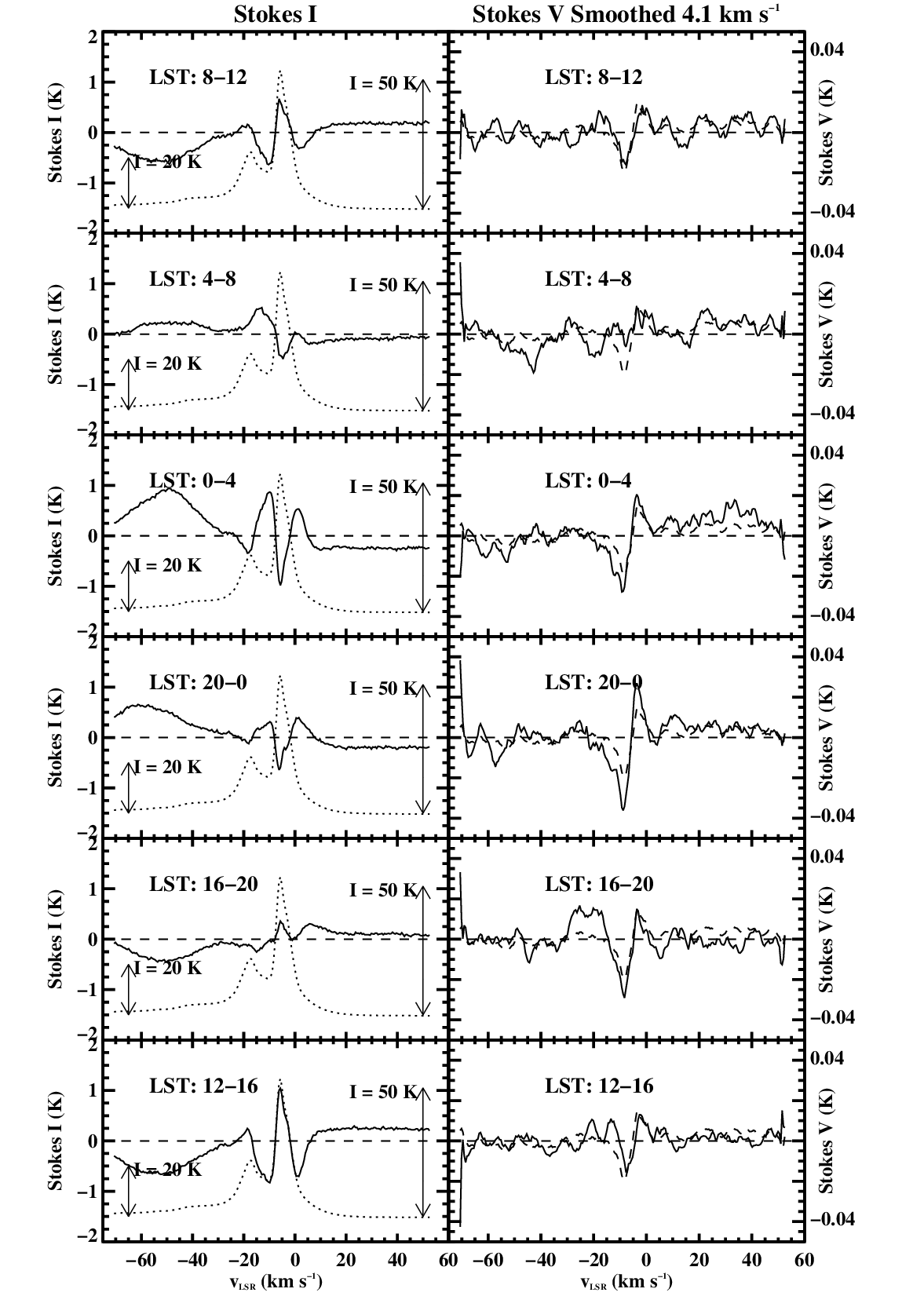}
  \end{center}
  \caption { Changes with time of the GBT 21 cm line profile for the North
    Celestial Pole (NCP) observed in 2003 September, with Stokes $I$ on the
    left and Stokes $V$ on the right. Each panel shows averages over 4
    hours of LST. For Stokes $I$, the solid-line profiles are differences
    between the 4-hour average and the 24-hour average. The dotted profile
    is the scaled-down 24-hour average Stokes $I$ profile; its left and
    right peaks have 20 and 50 K, respectively, as indicated. For Stokes
    $V$, the solid-line profiles are the observed ones and the dashed is
    the 24-hour average. \label{ncppaper} }
\end{figure}

Figure \ref{ncppaper} depicts the time variability of the GBT Stokes $I$
and $V$ 21 cm line profiles for observations of the North Celestial Pole
(NCP). For Stokes $I$, shown in the left panels, the data are averaged in
4-hour bins and the 24-hour average is subtracted. Differences typically
amount to $\sim$1 K For the peaks, amounting to a fractional change of a
few percent. More serious, however, at $-$40 km s$^{-1}$ the mean profile
has temperature $\sim$3 K and the fractional change is $\sim$30\%. The
profile distortions sometimes change rapidly with time. For example, the
$-$40 km s$^{-1}$ intensity peaks sharply in the {\rm LST} = 0--4 hr
bin.

For Stokes $V$, shown in the right panels, the data are averaged in 4-hour
bins and shown as solid lines; the 24-hour average is shown as dashed.
Changes are clearly obvious. Most extreme are the ${\rm LST} =6 \pm 2$ hr
bin, where the signal almost disappears, and the ${\rm LST}=18 \pm 2$ hr
bin, where there is a broad positive bump centered near $-$25 km s$^{-1}$.
Other bins also show shape changes.

\section{SIDELOBES RELATED TO THE SECONDARY REFLECTOR}
\label{sidelobeobsi}
\label{spillover}

\subsection{Descriptive Discussion}
\label{descriptivedisc}

\begin{figure}[!h]
  \begin{center}
    \includegraphics[width=6in]{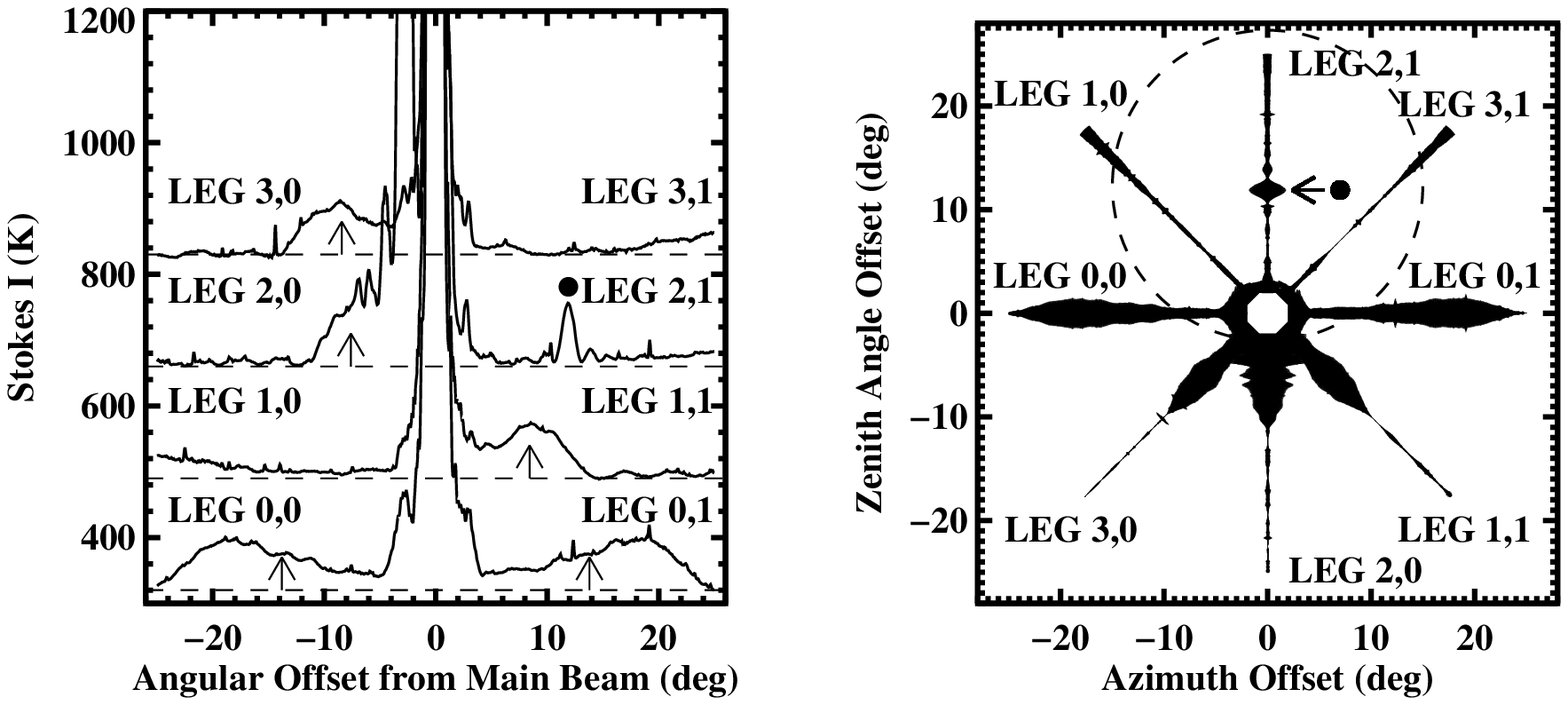}
  \end{center}
  \caption{Four Stokes $I$ scans (``Legs'') through the Sun in AZ/ZA
    great-circle offset coordinates.  In the left panel we plot the
    measured antenna temperatures versus the angular offsets from the main
    beam center. In the right panel, the azimuthal thicknesses of the lines are
    proportional to the measured temperatures; the dashed circle
    represents the approximate boundary of the secondary reflector as seen
    by the feed. The begin and end points of the legs are labelled on each
    panel by ``,0'' and ``,1'', respectively. On the left panel, the arrows
    lie $5^\circ$ outside the dashed circle. On both panels, the spot marks
    the Arago spot.
    \label{sunscansi}}
\end{figure}

\begin{figure}[!h]
  \begin{center}
    \includegraphics[width=5in]{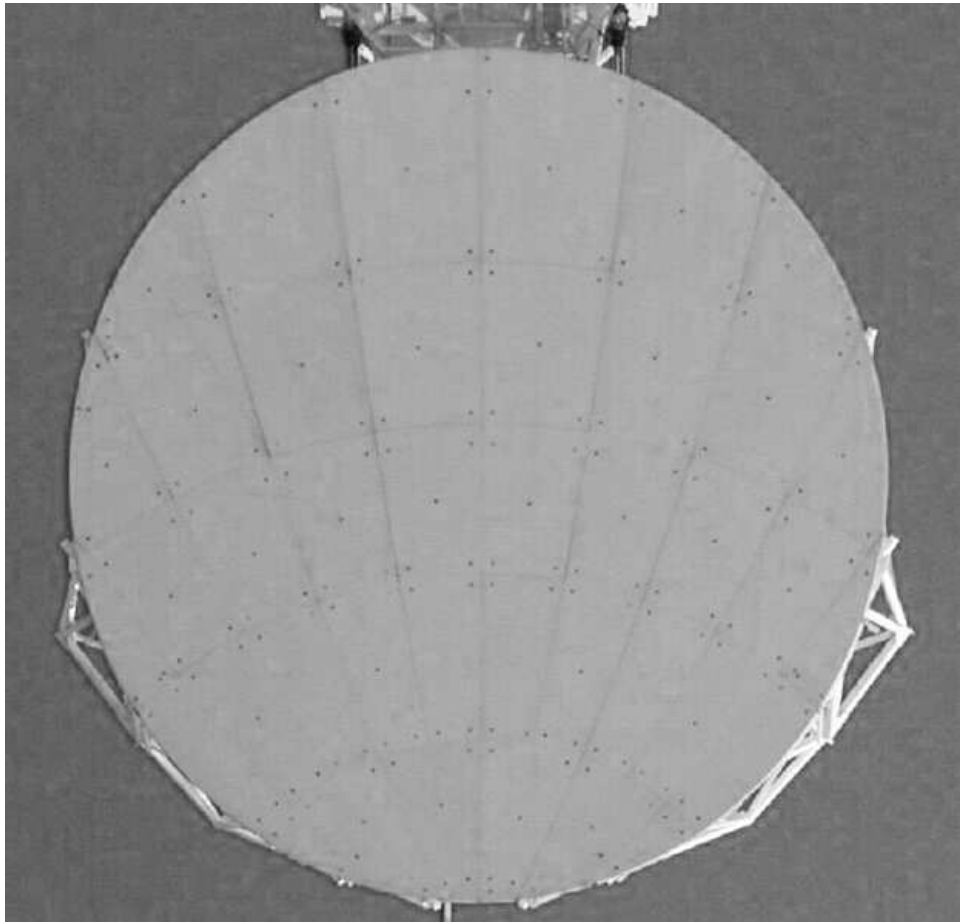}
  \end{center}
  \caption{The GBT secondary as seen from the feed, oriented so that the
    feed leg is on top---the same orientation as in Figure
    \ref{sunscansi}. The secondary looks slightly elliptical, with the
    major axis in the vertical (ZA) direction.  Image courtesy of NRAO/AUI.
    \label{gbtsec}}
\end{figure}

Figure \ref{sunscansi} exhibits Stokes $I$ versus AZ/ZA great-circle offset
coordinates for four scans through the Sun taken on 2003 September 19; data
on 2003 September 7 are essentially identical, so all features are
real. The right panel shows the four scans with azimuthal width
proportional to the observed intensity. The dashed circle shows the
approximate boundary of the secondary reflector as seen by the feed (the
actual boundary is slightly elliptical; see Figure \ref{gbtsec}). The left
panel shows the intensity profiles versus great-circle angular offset from
the main beam.

Figure \ref{sunscansi} shows that there is significant response outside of
the main beam. We represent this response by three components, which are
motivated by applying intuitive physical considerations to the reflector
geometries given by \citet{norrods96}: \begin{enumerate}

\item {\it The Spillover ring}. Leg 0 and the lower halves of Legs 1, 2,
  and 3 reveal the spillover contribution from the secondary, which peaks
  about $5^\circ$ outside of the dashed circle. The term ``spillover''
  refers to radiation transmitted from the feed toward the secondary
  reflector that would ``spill over'' its edge. Clearly, the spillover is
  not simply what we would find from geometrical optics, which would have a
  sharp cutoff located at the reflector edge.  Rather, the profiles show
  diffraction effects, with the peak lying outside the occulting edge and a
  gradual decrease in intensity toward the inside of the occulting area, as
  predicted by physical optics \citep{kraus88}.  Arrows on the left panel
  of Figure \ref{sunscansi} mark $5^\circ$ outside the reflector edge,
  roughly at the peak except for Leg 0. The decrease outside the peak is
  caused by the taper of the feed's illumination of the secondary.

  We represent this spillover pattern as being circularly symmetric about
  the axis of the subreflector center.  However, careful inspection of
  Figure \ref{sunscansi} reveals that this symmetrical condition does not,
  in fact, obtain. Comparing the peak locations with the arrows, which are
  located $5^\circ$ outside the subreflector edge, it is clear that the
  radial distance of the peaks of Leg 0 is larger than the radial distances
  of the other peaks.  A degree of noncircularity is expected because the
  secondary, as seen from the feed, is slightly elliptical; see Figure
  \ref{gbtsec}.  From this figure, we see that the secondary is slightly
  narrower in the AZ direction.  Because Leg 0 samples the AZ direction
  more than the ZA direction, the narrower ellipse projection suggests that
  the peaks along Leg 0 should lie {\it closer}, not further, than the
  other scans' peaks, opposite to what we see. We attribute this apparent
  turnabout to structural components that lie behind the secondary and
  protrude out from its projected edge, which are apparent as the ``ears''
  in Figure \ref{gbtsec}; Leg 0 passes close to these ears.
  
\item {\it The Arago spot}. A striking feature on Leg 2 is a small, intense
  spot, marked with a black spot. This is the famous Arago spot of an
  occulting disk.\footnote{As outlined in \citet{sommerfeld54}, this spot
    is a direct prediction of Fresnel's wave theory of light, which had
    been submitted for consideration in the 1819 Grand Prix of the
    Acad\'{e}mie des Sciences.  The committee included Poisson, Biot, and
    Laplace, and was chaired by Arago.  Poisson used Fresnel's theory to
    show that one would expect a bright spot anywhere along the central
    axis behind the blockage of a circular disk, then cited his own
    remarkable prediction as an objection to the wave theory given that no
    such spot is seen.  However, Arago actually performed the experiment
    and discovered that the spot did, in fact, exist!  Despite this
    experimental triumph, most contemporary physics texts use the term
    ``Poisson spot,'' while a few experimental texts prefer the term
    ``Arago spot.'' Being observers with a sympathetic eye to experiment,
    we prefer to honor Arago.}  According to \citet{sommerfeld54}, the
  intensity of the Arago spot for incoming parallel plane waves should be
  equal to the peak intensity of radiation near the edge of the
  blockage. Figure \ref{sunscansi} shows that this condition is achieved
  quite closely. We measure the Arago spot to be centered $0.39^\circ$
  below the subreflector center (i.e., toward the primary beam); this
  offset occurs because the axis of the secondary is tilted with respect to
  the line joining the feed and subreflector center.

\item {\it The Screen component}. Figure \ref{sunscansi} shows that there
  is excess intensity near the main beam, which we call the ``screen''
  component. Careful examination of Figure \ref{sunscansi} shows that this
  excess tends to be asymmetric about the main beam center, with excess
  intensities tending toward negative ZA offsets. This is particularly
  apparent when comparing the lower half of Leg 2 with those of Legs 1 and
  3, and also with the nearby structure of Leg 0. It seems like a separate
  beam a few degrees in size that is offset from the main beam center by a
  few degrees.

  In our mind's eye, we attribute this to the reflecting ``screen,'' which
  controls the spillover radiation that hits the structural components of
  the feed arm. At the top of Figure \ref{gbtsec} we see these feed-arm
  structural components. This jumble of metal beams would uncontrollably
  scatter the spillover radiation. To control this component, the telescope
  engineers installed a flat reflecting screen on the feed arm near the
  subreflector \citep{norrod90}.  This screen directs radiation coming from
  the feed back to the primary reflector, and then to cold sky.  It is as
  if a separate source of radiation centered on the screen illuminates the
  primary reflector; this source, the screen, is located near the
  intersection of Leg 2 with the dashed circle at the top of Figure
  \ref{gbtsec}, at positive ZA offset with respect to the subreflector.
  Being offset toward positive ZA offset, upon reflection from the primary
  reflector it would point toward negative ZA offset. We think that this
  might be responsible for the excess asymmetric intensity and this is why
  we call this the ``screen'' component.
  
  Radiation that is removed from the spillover by the screen, which we
  discussed in the above paragraph, eliminates radiation from an angular
  slice of the spillover pattern. We don't know how big this slice is, so
  we incorporated it as an unknown in our least-squares fits described
  below in \S \ref{lsfits}.

\end{enumerate}

\subsection{Caveat}

Our characterization of the GBT sidelobe structure is necessarily
approximate because we have limited mapping data in Figure
\ref{sunscansi}. Our characterization is based on this sparse sampling
together with physical reasoning and intuition. The quantitative
representation that we present below should be regarded as a convenient
approximation meant for the purposes of correcting 21 cm line data, not
as an accurate description of the sidelobes.

\subsection{Quantitative Representation}
\label{quantitativerep}

\begin{figure}[!h]
  \begin{center}
    \includegraphics[width=3in]{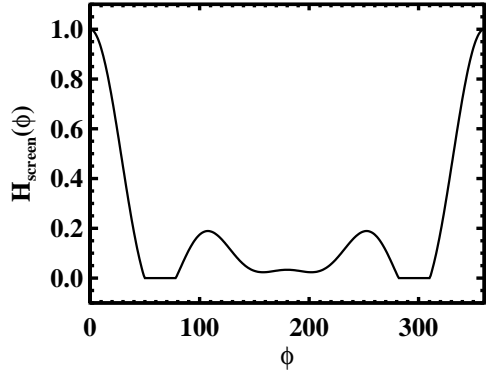}
  \end{center}
  \caption{$H_{\rm screen}(\phi)$ of equation (\ref{screeneqn}) versus $\phi$.
    \label{screenplot}}
\end{figure}

In the equations below, the intensity scale is parameterized by the
multiplying coefficient $H$. The units of $H$ are left undefined. Rather,
as explained in \S \ref{details}, we normalize the area of each component's
beam to unity so that its integrated response over the sky is in units of
the main beam's integrated response. For the three components, we have:
\begin{enumerate}

\item {\it The Spillover ring}. We represent the spillover ring by a sum of
  two two-dimensional Gaussian rings. The response $P_{\rm ring}(\theta)$ of
  each Gaussian ring is characterized by a radial distance $\theta_{\rm ring}$
  of its peak from the {\it center of the subreflector}, an angular
  half-power full width FWHM, and a height $H$.  That is, each ring has the
  functional form
  \begin{equation}
    P_{\rm ring}(\theta) = H_{\rm ring}
    \exp\left[ - \frac{ \left( \theta - \theta_{\rm ring}\right)^2 }
      { 2 ({\rm FWHM} /2.35)^{2} } \right]\,,
  \end{equation}
  where $\theta$ is the angular distance from the subreflector center. The
  parameters for the two Gaussians are: $H_{\rm ring} = [1.14, 1.0]$
  (arbitrary units); $\theta_{\rm ring} = [20.6^\circ, 16.4^\circ]$; ${\rm
    FWHM} = [3.3^\circ, 8.8^\circ]$. To account for the radiation scattered
  by the screen, we remove an angular slice of total width $49^{\circ}$
  centered on the feed leg---centered opposite the main beam as seen from
  the subreflector center.
  
\item {\it The Arago Spot}.  We represent the response of the Arago spot
  $P_{\rm spot}$ by a two-dimensional Gaussian spot whose center is offset
  $0.39^{\circ}$ toward the main beam from the subreflector center; its
  FWHM is $1.09^\circ$. The functional form is
  \begin{equation}
    P_{\rm spot}(\Delta {\rm AZ}, \Delta {\rm ZA}) = H_{\rm spot}
    \exp\left[  - \frac{ \Delta {\rm AZ}^2 + (\Delta {\rm ZA}+0.39)^2 }
      { 2 ({\rm FWHM}/2.35)^2 } \right]\,,
  \end{equation}
  where $(\Delta {\rm AZ}, \Delta {\rm ZA})$ are measured in degrees from
  the center of the subreflector as seen by the feed.

\item {\it The Screen component}. We represent the screen response as a
  two-dimensional Gaussian ring with ${\rm FWHM} = 0.5^\circ$ and radial
  offset from {\it main beam center} $\theta_{\rm screen}=2.8^{\circ}$. The
  amplitude of the Gaussian depends on $\phi$, the azimuthal angle around
  main beam center in $(\Delta {\rm AZ}, \Delta {\rm ZA})$ space.  The functional form
  is
  \begin{equation} \label{screeneqn}
    P_{\rm screen}(\theta, \phi) = H_{\rm screen}(\phi)
    \exp \left[ - \frac{ \left(\theta - \theta_{\rm screen}\right)^{2} }
        { 2({\rm FWHM}/2.35)^2 } \right]\,,
  \end{equation}
  where $\theta$ is the angular distance from main beam center and $\phi =
  \arctan( \Delta {\rm AZ}/\Delta {\rm ZA})$, measured counterclockwise from the
  $-\Delta {\rm ZA}$ axis. We estimated the amplitude $H_{\rm
    screen}(\phi)$ from the Sun scan legs seen in Figure \ref{sunscansi}:
  we forced an 8-term Fourier representation, which is shown in Figure
  \ref{screenplot} and is clearly a crude approximation.

\end{enumerate}

\subsection{Pictorial Representation}

\begin{figure}[!h]
  \begin{center}
    \includegraphics[width=6in]{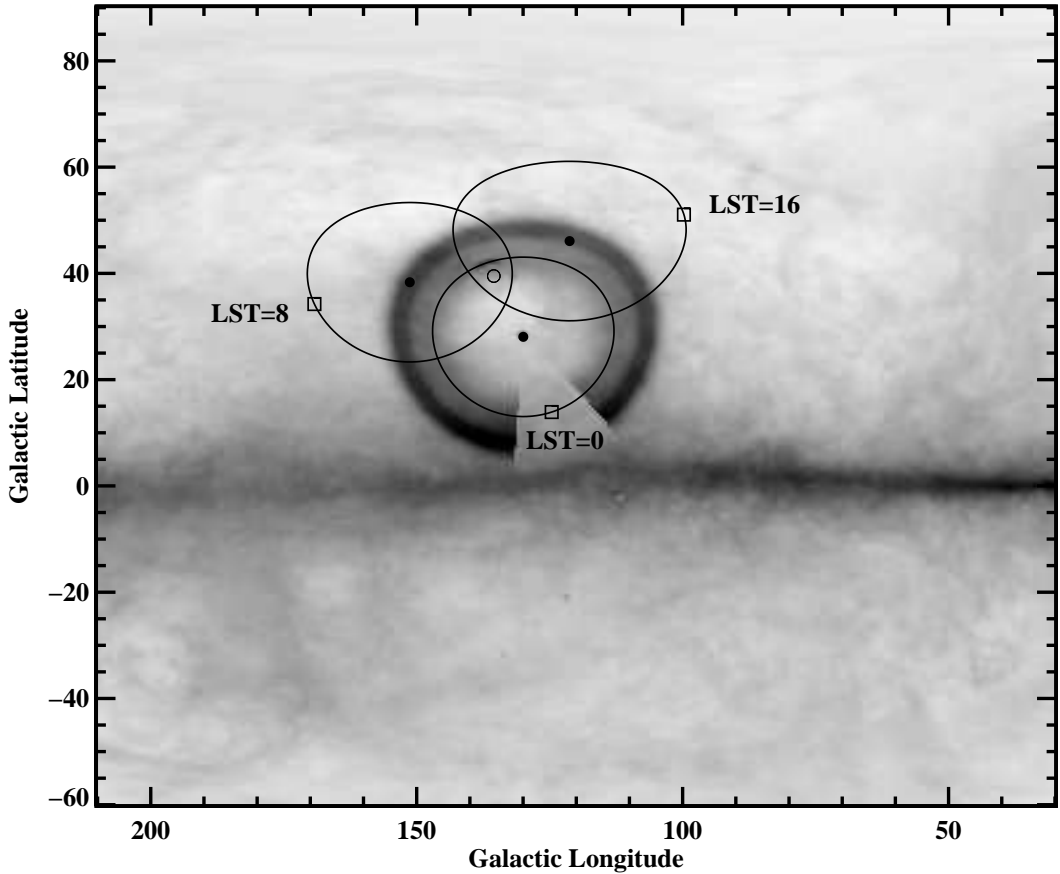}
  \end{center}
  \caption{Images of the secondary-reflector sidelobes at three LSTs for
    the position G135.5$+$39.5, superposed on the integrated 21 cm line
    intensity from the LAB survey. The empty circle is the source position.
    The filled circles are the Arago spot. The solid $30^{\circ}$-diameter
    circles approximately show the subreflector edge as seen by the feed.
    Each of the three squares lies on its subreflector circle just inside
    the feed arm.  The circular grayscale image has darkness proportional
    to the spillover pattern; this is shown only for LST=0.  The
    blanked-out portion of the spillover pattern is the angular slice of
    the screen.  \label{beamfig}}
\end{figure}

The discussion and equations of \S \ref{descriptivedisc} and \S
\ref{quantitativerep} represent an attempt to mathematically approximate
the contributions of the expected sidelobe response of the GBT. We can gain
an understanding of these sidelobes by mapping them on the sky.  Figure
\ref{beamfig} shows the spillover lobe, its screen slice, and the Arago
spot when observing the position G135.5$+$39.5 at LST = 0, 8, and 16 hr.
These are all evaluated for the adopted parameter values given in \S
\ref{lsresults}.

The spillover lobe intensity is shown by the large, circular grayscale
ring.  It peaks about $5^\circ$ outside of the subreflector edge, which
is shown as a $30^{\circ}$-diameter circle. The Arago spot lies about
halfway between the main beam (empty circle) and the feed arm (which
lies just outside the square). We don't show the screen component on
Figure \ref{beamfig}, because it lies close to the main
beam---$2.8^\circ$ away, on the opposite side of the main beam (i.e.,
the source position) from the Arago spot.

At LSTs near 1 hr, the spillover lobe brushes the Galactic plane; with the
large velocities of Galactic rotation, this produces significant
contributions to the observed profile at intermediate and high velocities.
12 hours later the spillover lies far from the Galactic plane and there is
no measurable contribution.

\section{ THE STOKES $I$ DIFFRACTION RINGS OF THE PRIMARY BEAM}
\label{diffraction}

\subsection{Descriptive Discussion}
\begin{figure}[!h]
  \begin{center}
    \includegraphics[width=6.0in]{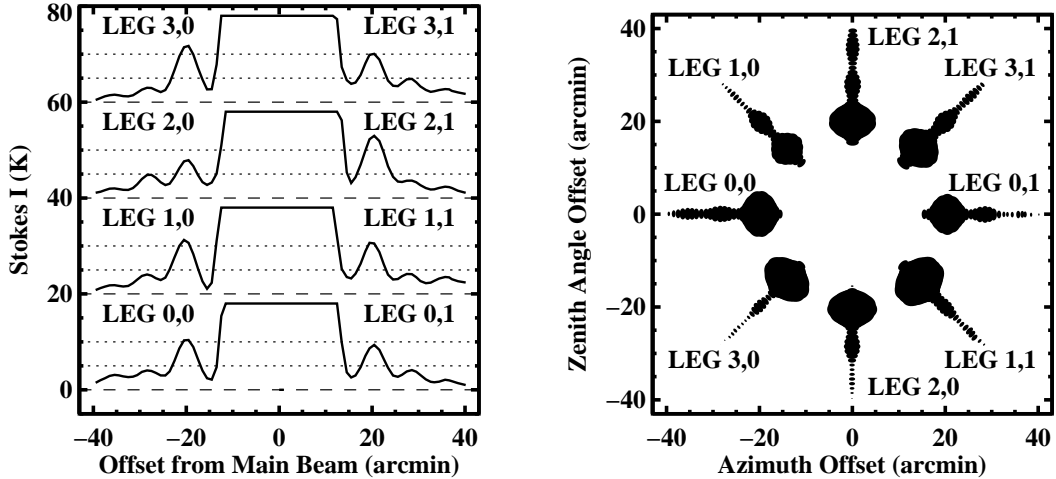}
  \end{center}
  \caption{Four Stokes $I$ scans (``Legs''; displaced zeros) through Cas A in AZ/ZA
    great-circle offset coordinates, showing the first three diffraction rings.
    In the left panel we plot the measured antenna temperatures versus the
    angular offsets from the main beam center. In the right panel, the
    azimuthal thicknesses of the lines are proportional to the measured
    temperatures. The begin and end points of the legs are labelled on each
    panel by ``,0'' and ``,1'', respectively. On the left, dotted lines show
    intensity offsets of 5 and 10 K for each leg.
    \label{casscansi}}
\end{figure}

Figure \ref{casscansi} represents the GBT diffraction rings as probed by
scans through Cas A.  The results are shown in two ways, with intensity
versus position offset for four observed legs in the left panel and a map
in the right.  The first ring is not circularly symmetric about beam
center. Rather, it is strongest at positive ZA offset and weakest at
negative offset. This is clear from examining Leg 2 on the left panel,
whose first ring peak at positive offset exceeds 10 K and at negative
offset is obviously less. In other words, the rings' responses depend on
$\phi$ [here, $\phi = \arctan(\Delta {\rm ZA}/\Delta {\rm AZ})$]. In
contrast, the $\phi$ asymmetry for the second ring looks opposite to that
of the first. Clearly, the situation regarding asymmetry in $\phi$ is
complicated and cannot be easily described analytically. We use an
oversimplified quantitative model for the diffraction rings because their
structure is complicated and we do not have complete and accurate data.

We split the diffraction rings into two parts: (1) the average over $\phi$
and (2) the portion that changes with $\phi$. (If we were to represent the
$\phi$ behavior with a Fourier series, the first would be the ``DC''
term): \begin{enumerate}

\item For the $\phi$-averaged part, the 21 cm line profile contribution is
  independent of parallactic angle PA (i.e., hour angle). This contribution
  is small. The peak amplitude of the first ring is about 1000 times weaker
  than the peak of the main beam. The angular areas of the rings exceed
  those of the main beam. With these, the net angle-integrated beam
  response of the rings is no more than $\sim$1\% of the main beam's.
  Because the rings lie close to the main beam, the spectral shape of their
  contribution is similar to that of the main beam's, so they simply
  augment the main beam's profile by a small amount. Our analysis lumps
  these together, so that elsewhere in this paper when we use language such
  as the ``integrated response of the main beam,'' we in fact mean the
  contribution of the main beam plus the $\phi$-averaged response of the
  diffraction rings.

\item For the $\phi$-variable portion, we adopt a very simple model. We
  assume (1) that the $\phi$ dependence is independent of radial offset
  from beam center, and (2) that the $\phi$ dependence is described by the
  lowest Fourier mode [i.e., $\propto \cos(\phi - \phi_0)$]. The first
  assumption is clearly invalid, given our discussion of Figure
  \ref{casscansi} in the above paragraph, but as we mentioned, the
  situation is complicated. We expect our oversimplified ring treatment to
  produce inaccuracies in our predicted spectra.
\end{enumerate}

\subsection{The ``Nearin'' Component}

With our assumptions, we can represent the $\phi$-variable portion by the
difference between two primary beams pointing on opposite sides of beam
center. We name this the ``Nearin'' component. We don't use the name
``Ring,'' because this representation is only an approximation.  To
determine the appropriate parameters, we assumed the position difference
between the beams to be 5 arcmin and performed the least-squares procedure
of \S \ref{lsfits} for a grid of $\phi_s$ values, where $\phi_s$ is the
angle of the position difference; we obtained $\phi_s = 180^\circ$. (If the
first ring alone were responsible, the left panel of Figure \ref{casscansi}
shows that we would have obtained $\phi_s = 90^\circ$.)

The other three sidelobe contributions have nonzero (positive)
sky-integrated beam responses. However, the sky-integrated response of
the Nearin component is zero, because it is the difference between two
main beam pointings. In particular, the profile predicted from this
component can be both positive and negative. Thus, our discussion below
about the physical meaning of the derived $\alpha$ coefficients in \S
\ref{details} applies to the other three sidelobe contributions, but not
to the Nearin one.

\section{ DERIVING THE STOKES $I$ SIDELOBE AMPLITUDES USING 21 CM
  LINE DATA AND THE METHOD OF LEAST SQUARES}
\label{lsfits}

\subsection{The General Approach}

While our Sun scans of \S \ref{sidelobeobsi} allow us to measure the
sidelobe beam {\it shapes} with fair accuracy, we cannot measure the {\it
  amplitudes}. The peak response of these sidelobes is more than $\sim$1000
times smaller than the peak on-axis beam response (for the diffraction
rings) and at least another 10 times smaller for the spillover and spot
beams. The linear dynamic range of modern radioastronomical receiving
systems is limited by both analog and digital electronics and the net
response is highly nonlinear over such large ranges. Therefore, rather than
try to correct for saturation effects by bootstrap calibration techniques
or to rely on theoretical calculations, we use observations of the 21 cm
line at four positions to define the approximate contaminating beam shapes
and use least-squares fitting to obtain their amplitudes. We present our
technique below, but first we mention a set of details that should be kept
in mind.

\subsection{Four Important Details} \label{details}

There are four important details that should be considered concerning these
fits:
\begin{enumerate}

\item The spillover and spot cover areas of sky located at considerable
  angular distances from the observed position. Doppler corrections differ
  substantially over these angular distances. The LAB survey provides
  profiles in the Local Standard of Rest (LSR) system; given that we
  observe in the topocentric reference frame, we must convert all LAB
  profiles to topocentric velocities.  Upon obtaining the estimated
  sidelobe response, the spectrum must be converted back to the LSR
  reference frame for the position actually pointed at.

  \begin{deluxetable}{cc}
    \tablewidth{0pc} \tablecaption{LAB/GBT Intensity Ratios
      \label{intratio} } \tablehead{ \colhead{ Source Name }& \colhead{
        (LAB/GBT) }} \startdata
    G139.1$+$0.7 & 0.845  \\
    W3off (G133.8$+$0.8) & 0.804  \\
    G135.5$+$39.5 & 1.409  \\
    NCP (G122.9$+$27.1) & 1.275 \enddata
  \end{deluxetable}

\item For each of the four positions, the several hundred spectra included
  in each fit were obtained over at least a 24-hour interval and often over
  several days. We find that the gains are not perfectly calibrated. That
  is, even though the spectral {\it shapes} are unaffected, the calibrated
  {\it intensity scales} change with time as these gains change.  We need
  to assign each observed spectrum its own gain and include the set of
  gains as unknowns in the fit.

\item The overall intensity scale of the GBT spectra, relative to that of
  the LAB spectra, changes from one position to another as shown in Table
  \ref{intratio}.  The ratio of LAB to GBT intensities (LAB/GBT) ranges
  from 0.80 to 1.41, changing by a factor of 1.75, with smaller values
  associated with sources in the Galactic plane and larger values
  associated with high-latitude positions. These numbers are obtained from
  the least-squares fits for $G_s^*$ in equation (\ref{eqnf0}), so they are
  not a result of sidelobe contributions.

  The sense of this---higher ratios in the Galactic plane---is contrary to
  what we'd expect from the angular structure of the \ion{H}{1} line. In
  the Galactic plane, the \ion{H}{1} line intensity has a relatively small
  fractional variation over the scale of 40 arcminutes, so the \ion{H}{1}
  emission fills the main beams of both the GBT and the LAB telescopes. In
  contrast, for the two high-latitude positions, the \ion{H}{1} structure
  is filamentary with widths of order $\le$10 arcmin, as shown by 100
  $\mu$m IRAS maps \citep{jacksonwg03} and, also, our currently unpublished
  GBT 21 cm maps of this area. The GBT main beam is filled, or nearly
  filled, by the \ion{H}{1} emission; but the larger beams of the LAB
  telescopes are not filled---nor are their nearin diffraction lobes. So we
  expect {\it lower} LAB/GBT ratios for the high-latitude positions,
  contrary to the numbers in Table \ref{intratio}.

  We suspect that the suggested trend in Table \ref{intratio} for higher
  ratios at the high-latitude positions is an illusionary result of
  small-number statistics.  We specifically do not believe that varying
  ratios of LAB/GBT are produced by problems with the LAB spectra, because
  great care has been taken to ensure their accuracy and, also, they have
  been carefully corrected for sidelobe problems.  Therefore, we conclude
  that the intensity scale of the GBT \ion{H}{1} spectra can vary by
  factors in the above range, namely $\sim \pm 30\%$, for reasons that we
  cannot fathom.

\item The observed spectrum is contaminated by contributions from the
  spillover, the Arago spot, the screen, and the Nearin response. The first
  three of these have nonzero (positive) sky-integrated beam responses, and
  the Nearin one has zero integrated response. The following discussion
  applies only to the first three, not to the Nearin component.

  In calculating the beams associated with these three contaminants, we
  normalize each one so that its gain-area product is unity; we then
  simultaneously fit for the proper amplitude coefficients (and other
  quantities; see \S \ref{lstech}), which we call $\alpha$ with appropriate
  subscripts. With the normalized integrated beam response, $\alpha$
  represents the contaminating beam's integrated ``beam efficiency.'' That
  is, if the contaminating beam were placed in a blackbody cavity at
  temperature $T_{\rm bb}$, it would contribute $\alpha T_{\rm bb}$ to the
  observed antenna temperature $T_a$. Similarly for the main beam, we can
  define an amplitude coefficient $\alpha_{\rm mb}$, which would equal the
  classically-defined main-beam efficiency. That is, in the blackbody
  cavity the main-beam antenna temperature would equal $\alpha_{\rm
    mb}T_{\rm bb}$.

  We assume that the LAB survey temperature scale is absolutely correct in
  the sense that the spectral temperatures are, in fact, the actual
  brightness temperatures. Our least-squares solutions in \S \ref{lstech}
  force the GBT temperature scale for the ``true spectrum'' $T^*_c$ (see \S
  \ref{lstech}) to equal the LAB temperature scale. In essence, this is
  forcing $\alpha_{\rm mb}$ to equal unity. Thus, for one of these first
  three contaminating beams, its value of $\alpha$ represents its beam
  efficiency with respect to the main beam efficiency, i.e., $\alpha$ is
  equal to the contaminating beam's area-integrated response divided by
  that of the main beam.

  This is a somewhat awkward situation, because the sum of the four
  $\alpha$ values exceeds unity. That is, when the GBT is placed in a
  blackbody cavity of temperature $T_{\rm bb}$, the antenna temperature
  exceeds $T_{\rm bb}$ by a factor equal to the sum of all four $\alpha$
  values. However, if the contaminating beams do not see the walls of the
  cavity, then the antenna temperature is equal to the brightness
  temperature. This awkwardness pervades measurements of extended sources
  by all single dishes, unless the sidelobes and main beam responses are
  reconciled as they are for the LAB survey (see \citealt{hartmannkbm96}
  and \citealt{kalberlabhabmp05}). The proper reconciliation of these
  matters is a {\it fundamentally important contribution} by the LAB
  survey. The GBT, with its seeming inability to reliably and
  self-consistently measure \ion{H}{1} line intensities, cannot yet be
  regarded as a suitable system for obtaining accurately-calibrated
  \ion{H}{1} line profiles.

\end{enumerate}

\subsection{The Least-Squares Technique}
\label{lstech}

Suppose we have a number of spectra $S$, each with $C$ spectral
channels, taken over a mostly complete range of 24 hours in LST. Let the
subscripts $s$ and $c$ denote the spectrum number and spectral
channel number, respectively. Let the superscript $o$ designate the
observed value and $*$ the true value (which is the same as the noise-free
fitted value) that would be observed in the absence of sidelobe
contributions and gain errors. Let $T$ denote the intensity, which is
normally expressed in units of temperature. As we mentioned in \S
\ref{details}, the intensity scaling of each spectrum (the ``gain''
$G_{s}$) is not perfectly calibrated so we solve for the set of spectral
gains. As we discussed in \S \ref{details}, we normalize each
contaminating beam so that its gain-area product is unity; then we fit
for the proper scaling parameter, which we call $\alpha$ with an
appropriate subscript.

The observed spectrum $T^{o}_{c,s}$ is equal to the true spectrum $T_c^*$
plus the contaminating contributions from the spillover, the Arago spot,
the screen, and the Nearin. We calculate the contaminating spectral
contribution of each of the first three by multiplying its beam by
appropriately velocity-shifted spectra from the LAB survey.  We can use the
LAB survey for these beams because their angular scales are large compared
to the LAB's 36-arcmin angular resolution.

However, the Nearin contribution depends on the angular structure close to
the GBT main beam, which is on a smaller scale than the LAB survey
resolution. For the Nearin contribution, we estimate the angular
derivatives of the emission brightness temperature from a 17-point sampling
with the GBT centered on the position observed, using the same technique
(the ``Z16 pattern'': i.e., 16 off-source positions and one on-source) as
\citet[][\S 2]{heilest03i}.

The following equation expresses the observed spectrum $T^{o}_{c,s}$ in
terms of the true spectrum $T^{*}_{c}$ and the four contaminants
$T^{j}_{c,s}$, where the superscript $j$ runs from 0 to 3 and represents
spillover, spot, screen, and Nearin, respectively; $j$ also applies to
the beams' $\alpha$ coefficients as a subscript. Note that the
contribution from each contaminating spectrum changes with time, so it's
a function of spectrum number $s$.

\begin{equation} \label{eqnf0}
  T^{o}_{c,s}= G^{*}_{s} \left[ T^{*}_{c} +
    \sum_{j=0}^{j=3} T^{j}_{c,s} \alpha^{*}_{j} \right]\,.
\end{equation}

The total number of equations is $CS$. The unknowns are the starred
values: the $S$ values of gain $G_s^*$, the C values of the true
spectral channels $T^{*}_{c}$, and the four $\alpha_j^*$ values; the total
number of unknowns is $(C + S + 4)$. We have $CS$ equations, which is many
more than the number of unknowns.  This is a well-posed least-squares
problem.

It is also a {\it nonlinear} least-squares problem because the unknown
gains $G^{*}_{s}$ multiply the other unknowns. We linearize this equation by
Taylor expansion, setting each unknown equal to a ``guessed'' value with
subscript $g$ plus a correction preceded by $\delta$. For example, the true
value $T^{*}_{c} = T^{g}_{c} + \delta T_{c}$. In formulating the
equations of condition, we treat the guessed values as known, so the
corrections become the quantities to solve for. Having solved for the
corrections, we apply them to the guessed values and do the fit again,
iterating until convergence.

In terms of the guessed and correction values, equation (\ref{eqnf0})
becomes
\begin{equation}
   T^{o}_{c,s}= (G^{g}_{s} + \delta G_{s}) \left[
    (T^{g}_{c} + \delta T_{c}) + 
    \sum_j T^{j}_{c,s} (\alpha^{g}_{j} + \delta \alpha_j) \right]\,.
\end{equation}
The summed quantity $\sum_j T^{j}_{c,s} \alpha^{g}_{j}$, which consists of
the guessed values for the unknown coefficients, is the guessed value of
the observed spectral channel in each iteration, so it is natural to define
the correction to the observed value as
\begin{equation}
  \delta T^{o}_{c,s} =  T^{o}_{c,s} - T^{o,g}_{c,s}\,.
\end{equation}
Grouping and retaining only first-order terms, our equations of condition
become
\begin{equation} \label{eqcond0}
  \delta T^{o}_{c,s} = G^{g}_{s} \ \delta T_{c} +
    \left[ \sum_j G^{g}_{s} T^{j}_{c,s}\right] \delta \alpha_{j} +
    \left[ T^{g}_{c} + \sum_j T^{j}_{c,s}\alpha^{g}_{j} \right] \delta G_{s}\,.
\end{equation}
In these equations (of which there are a total number $CS$), the three
$\delta$ quantities represent the $(C+S+4)$ unknowns and all other
quantities are either measured or guessed, so are known quantities as far
as the fit is concerned.

There is one additional constraint we need to add as a final equation of
condition. Consider equation (\ref{eqcond0}). The last term of this equation
contains five temperature terms: the first, $T^{g}_{c}$, and the other
four, which are the sidelobe contributions $T^{j}_{c,s}\alpha^{g}_{j}$.
The first is much larger than the latter four. If we were to add a single
constant to {\em all} $\delta G_{s}$, and simultaneously scale down
$T^{g}_{c}$, the fit would be almost unchanged.  Thus, there is a
near-degeneracy between these quantities. If we allow this constant to
exist, then there is a high negative covariance (a near-degeneracy) between
the final values of (1) this constant value added to all $G^{*}_{s}$ and
(2) the amplitude scaling of $T^{*}_{c}$. We can remove this
covariance---and retain the mean of the original amplitude scaling---by
requiring the sum of the guessed gains to remain constant. This is
equivalent to requiring
\begin{equation} \label{eqcond1}
  \sum_{s=0}^{S-1} \delta G_{s} = 0\,.
\end{equation}
It is straightforward to form a matrix (the Numerical Recipes ``design
matrix''; \citealt[][\S 15.4]{presstvf92}) from the $CS$ equations
\ref{eqcond0} and the additional equation (\ref{eqcond1}). We then perform
the least-squares solution in the usual way.

\subsection{Solving for the Slice Angle}
\label{lsslice}

In \S \ref{lstech} above we do not mention the unknown angle of the
slice that is removed by the screen. Nevertheless, we fit for it as an
unknown parameter, in addition to the others mentioned above. Instead of
incorporating it into the equations of condition, however, we used the
``brute-force'' technique. That is, we selected a grid of values for the
slice angle and carried through the least-squares fit of \S \ref{lstech}
for each assumed value. Then we plotted the variance versus the assumed
slice angles. The minimum variance defines the fitted slice angle.

\section{LEAST-SQUARES FIT RESULTS FOR THE FOUR SOURCES}
\label{lsresults}

\subsection{Systematic Changes of Calibrated System Gains}
The system gain is obtained by turning on a noise diode, which provides a
calibrated broadband noise signal of known equivalent temperature. If this
calibrated noise is constant and the system is linear, then this procedure
reliably establishes the intensity scale for the measured 21 cm line
profiles. The true intensity scale must be obtained by accounting for the
atmospheric opacity, which we calculated using the procedure given by
\citet{hartmann94}. Although the opacity is small for the 21 cm line,
about 0.013 at the zenith, it increases with the zenith angle and would
appear as a time-variable gain as we track a source. We do, in fact, see
time-variable gains---even after removing the atmospheric opacity.

In the following discussion, the system gain $G_s$ is a normalized
version of that in equation (\ref{eqnf0}) above: for each source we normalize all the
gains so that their average is unity. With this, if the observed
calibrated spectral intensities are too high by a factor of 1.2, then the
gain $G_s$ is equal to 1.2. We also define the ``system
temperature'' $T_{sys,s}$ as the off-line system temperature for each
spectrum. As above in \S\ref{lsfits}, the subscripts $c$ and $s$ represent
spectral channel and spectrum number, respectively. We calculate these
quantities using a least-squares fit, as follows.

For each source, we begin by averaging all the spectra to obtain the
average spectrum $\langle T_{c}\rangle$.  For each spectrum, we obtain the
gain $G_s$ by comparing its 21 cm line intensity to that of $\langle
T_{c}\rangle$, and we obtain the system temperature $T_{sys,s}$ by
comparing the off-line intensities. Operationally, we obtain these
parameters by a linear least-squares fit for each spectrum in which the
independent variable is the average spectrum. That is, for each spectrum we
fit for the zero intercept $Z_{s}$ and the gain $G_s$ in the set of $C$
equations

\begin{equation}
T_{c,s} = Z_s + G_s \langle T_{c}\rangle\,.
\end{equation}

\begin{figure}[!p]
  \begin{center}
    \includegraphics[width=5.2in]{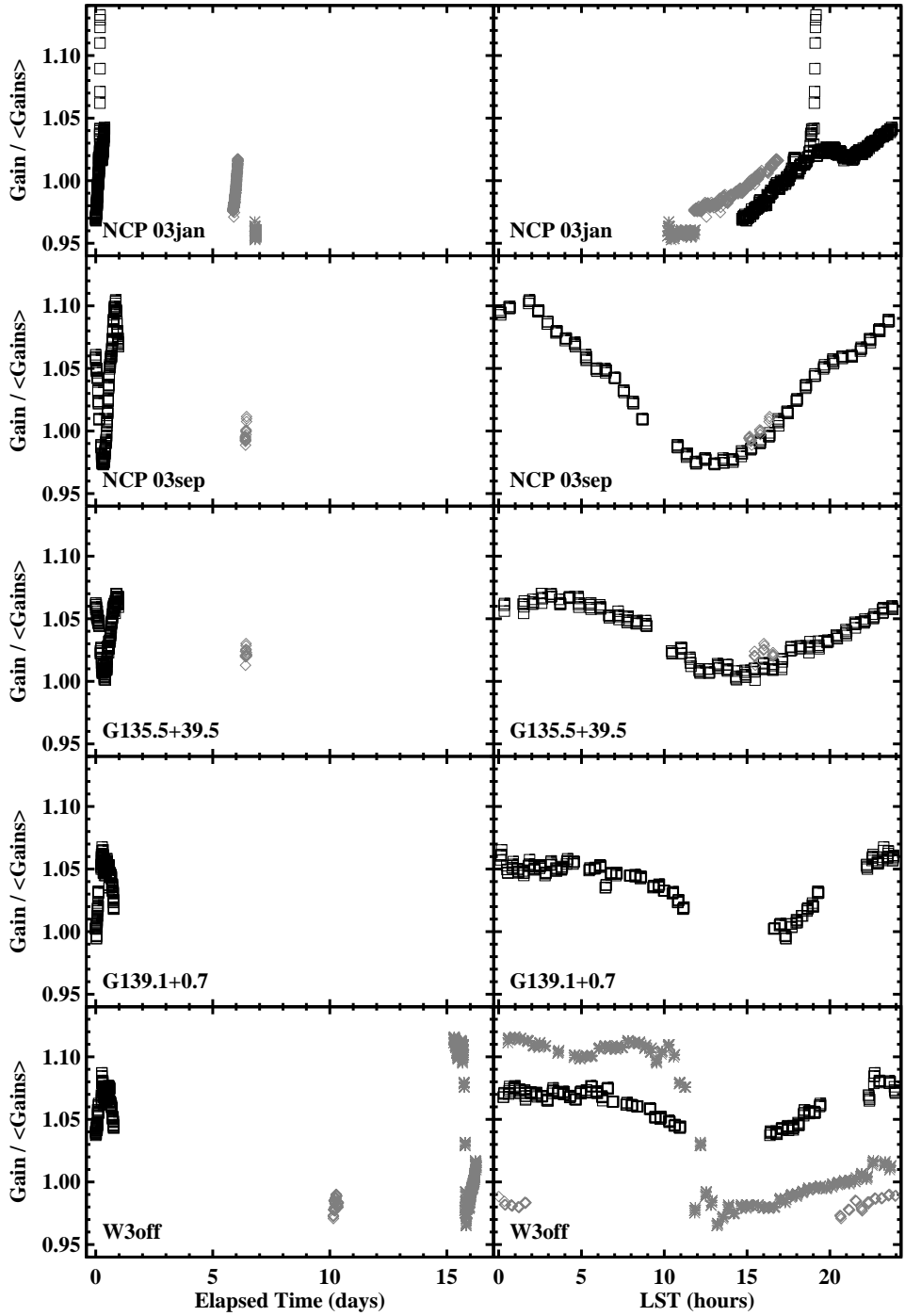}
   \end{center}
   \caption { Gains versus time for five data sets. The left-hand column
     shows elapsed time in days from the first observation for each
     data set. The right-hand column shows Local Sidereal Time (LST) in
     hours. Different days have different symbols. The top two rows are the
     North Celestial Pole (NCP) in 2003 January and 2003 September,
     respectively. G135.5$+$39.5 was observed in 2003 September and the other two,
     G139.1$+$0.7 and W3off, were observed in 2003 August.
     \label{allgains} }
\end{figure}

Ideally, all $G_s$ should be unity to within the uncertainties.
Unfortunately, this is not the case. The left-hand panels of Figure
\ref{allgains} show the gains for five sets of observations versus elapsed
time in days from the first observation (in addition to the four sources
above, we include 2003 January observations of the NCP). Variations range
up to $\sim$10\% displaying a systematic behavior with time. Since these
variations are not random, we have examined the systematic behavior and
find no clear dependence that applies to all sources. However, there is a
tendency for the gains to change in a somewhat common fashion with LST.
The right-hand panels of Figure \ref{allgains} show the gains $G_s$ versus
LST. For LSTs below about 13 hours, the gains tend to decrease with time;
above 13 hours, they tend to increase with time.

The data for the top two rows of panels, the NCP in 2003 January and the
NCP in 2003 September, are particularly intriguing. The telescope doesn't
move while tracking the NCP, so one cannot attribute the changing gains to
a changing telescope position.  Moreover, one cannot attribute it to the
influence of the Sun, because the right ascension of the Sun is 8 hours
different for the two data sets. It seems that the changing gains must be
produced internally to the receiver electronics; the most straightforward
explanation would be a changing noise diode temperature, but we are at a
loss to propose an explanation of why it should change systematically with
LST. Of course, we have limited data, so perhaps the systematic change with
LST is only an illusion.

\begin{figure}[!p]
  \begin{center}
    \includegraphics[width=6in]{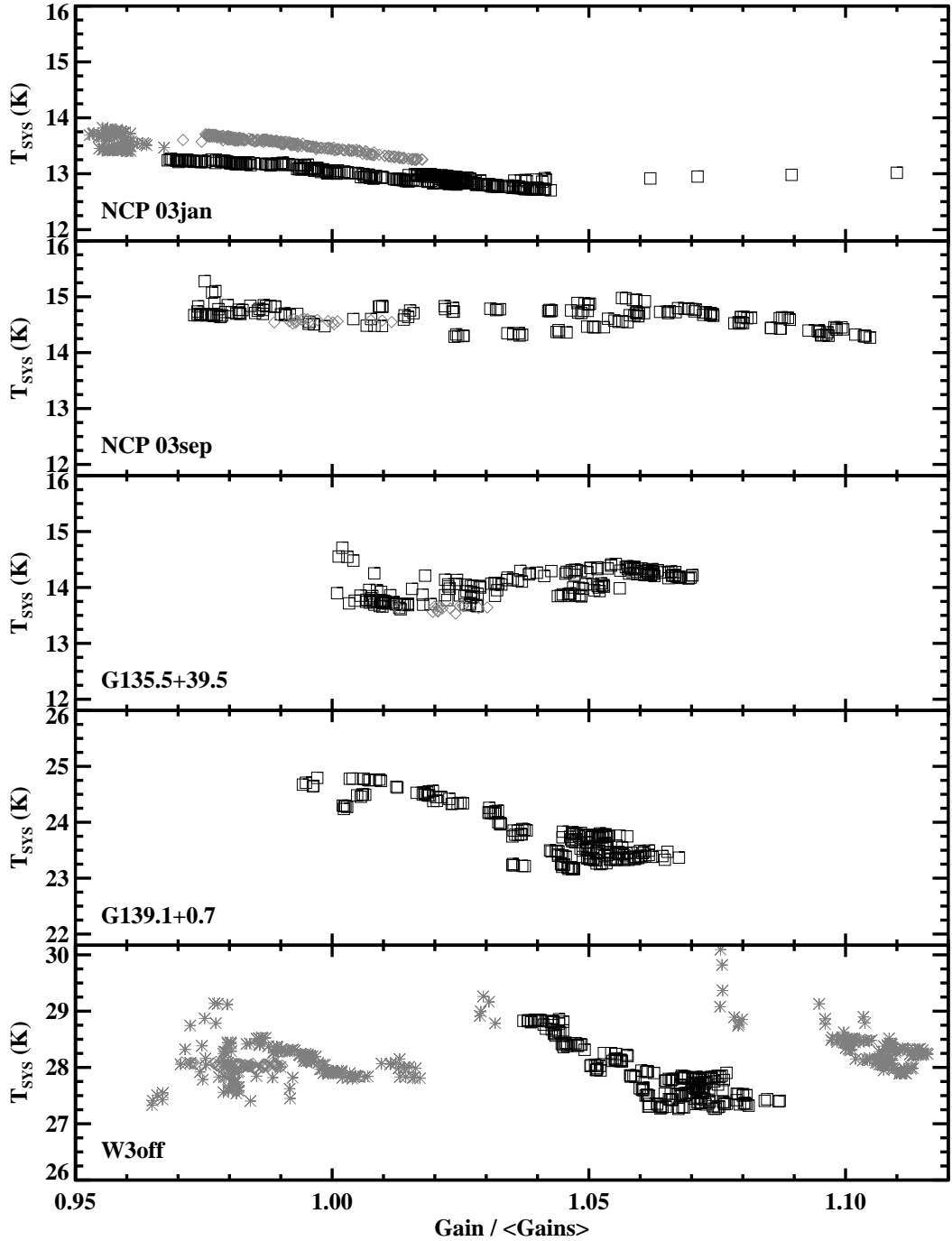}
   \end{center}
   \caption {System temperatures versus the normalized gains of Figure
     \ref{allgains}. Different days have different symbols, the same as
     in Figure \ref{allgains}.
     \label{hilatgains} }
\end{figure}

Figure \ref{hilatgains} shows that, for some sources, the system
temperature $T_{sys,s}$ varies systematically with the gain. The top
panel (NCP in 2003 January) and the bottom two panels exhibit a very
clear inverse correlation. In contrast, the NCP in 2003 September shows
no visual dependence and G135.5$+$39.5 exhibits a small direct, instead of
inverse, correlation. We can think of no reason why there should be any
relation between system gain and temperature at the levels which seem so
visually obvious in Figure \ref{hilatgains}.

The important point for measurement of accurate 21 cm line profiles,
however, is that the gains change substantially with time. This means
that the intensity scales of 21 cm profiles obtained with the GBT
cannot be trusted at these levels.

\subsection{The Derived Coefficients and Their Grand Averages }
\label{grandavg} 
\begin{deluxetable}{cccccc} 
  \tablewidth{0pc}
  \tablecaption{Derived Parameters \label{lscoeffs} }
  \tablehead{
    \colhead{ Source Name }&
    \colhead{ slice angle } &
    \colhead{ $\alpha_{\rm spillover}$ }&
  \colhead{ $\alpha_{\rm spot}$ } &
  \colhead{ $\alpha_{\rm screen}$ } &
  \colhead{ $\alpha_{\rm nearin}$ }  }
  \startdata
  G139.1$+$0.7         & $44.3^{\circ}$  & 0.089 & 0.0012 & 0.0078  &  0.023 \\
  W3off (G133.8$+$0.8) & $49.7^{\circ}$  & 0.085 & 0.0007 & 0.0062  &  0.021 \\
  G135.5$+$39.5        & $61.1^{\circ}$  & 0.092 & 0.0013 & $-$0.0241 &  0.012 \\
  NCP (G122.9$+$27.1)  & $46.6^{\circ}$  & 0.073 & $-$0.0008& 0.0034  &  0.012 \\
  ADOPTED            & $49^{\circ}$  & 0.087 & 0.0010 & 0.0070  &  0.022 \enddata

\end{deluxetable}

In our least-squares fit for each position, we derive the true profile $
T^{*}_{c}$ and a coefficient $\alpha$ for each of the four stray beams. We
do this for a set of assumed slice angles, as explained in \S
\ref{lsslice}.  Table \ref{lsresults} presents these fitted coefficients
for each of the four positions. We don't list the formal uncertainties,
which are meaninglessly small because even though there is a huge number of
independent points in the fit (e.g., for 200 spectra with 400 channels each
there are 80000 data) there is a large correlation among spectra taken
nearby in time. Rather, the coefficient uncertainties are better estimated
from the variation among derived values for the four positions together
with consideration of each coefficient's contribution to each source's
profile.  The last line in the table lists our adopted coefficient values.

The largest contaminating beam area is the spillover, whose integrated
beam response $\alpha_{spillover}$ is almost $10\%$ of the main
beam's. The screen's integrated beam response is smaller by a factor of
about 12 and the spot's response is even smaller. The spot and the
spillover cover fairly similar regions in realistic situations, so
usually the spot's contribution to the observed 21 cm line profile
should be unimportant. This is not necessarily true for the screen,
however, because it lies close to the main beam and the 21 cm line
intensity here might be much higher than that covered by the
spillover---or vice-versa.

In principle, conservation of energy requires that the power removed
from the spillover by the screen should appear in the screen
component. That is, with a slice angle of $49^\circ$, we should have
\begin{equation}
  \frac{49}{360} = \frac{ \alpha_{\rm screen} }
  { \alpha_{\rm spillover}+\alpha_{\rm screen} } .
\end{equation}
The ratio on the left is 0.14 and that on the right is 0.075, so this
relationship is not satisfied at the factor-of-two level. This probably
means that we have not properly estimated the properties of the screen
component.

\subsection{Fit Results for Each Source}
\label{fitresults}

\begin{figure}[!p]
  \begin{center}
    \includegraphics[width=5in]{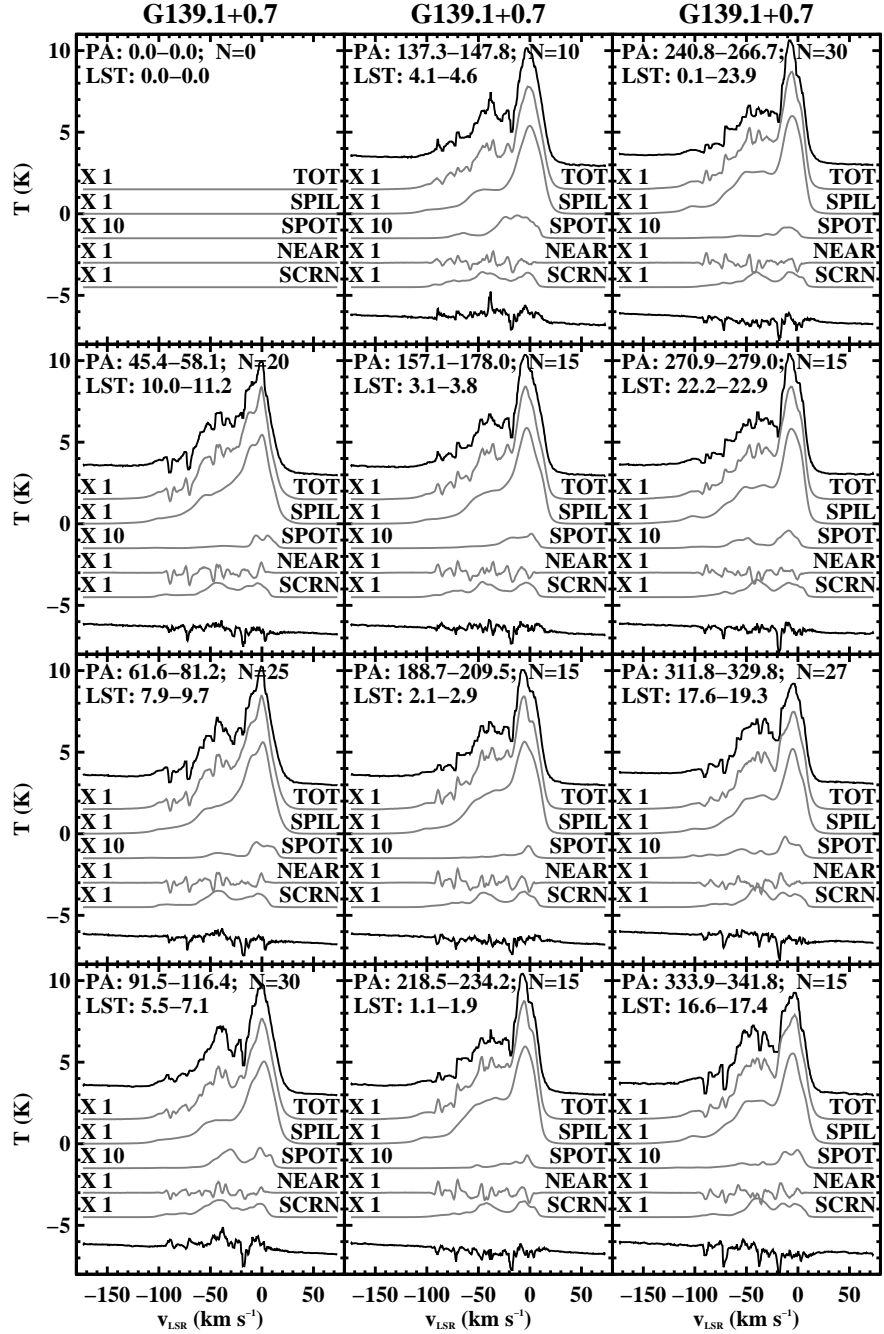}
  \end{center}
  \caption {Observed and predicted sidelobe contributions for
    G139.1$+$0.7.  Each panel shows the averages for $30^{\circ}$-wide
    bins of parallactic angle PA, with successive bins running down the
    page and then to the next column. The top curve (in black) in each
    panel is the total observed sidelobe contribution; the next curves,
    in gray, are the predicted sidelobe contributions; the bottom
    curve, in black, is the difference between the total observed and
    predicted contributions. For more details, see text (\S
    \ref{fitresults}).
    \label{g139fit} }
\end{figure}

\begin{figure}[!p]
  \begin{center}
    \includegraphics[width=5in]{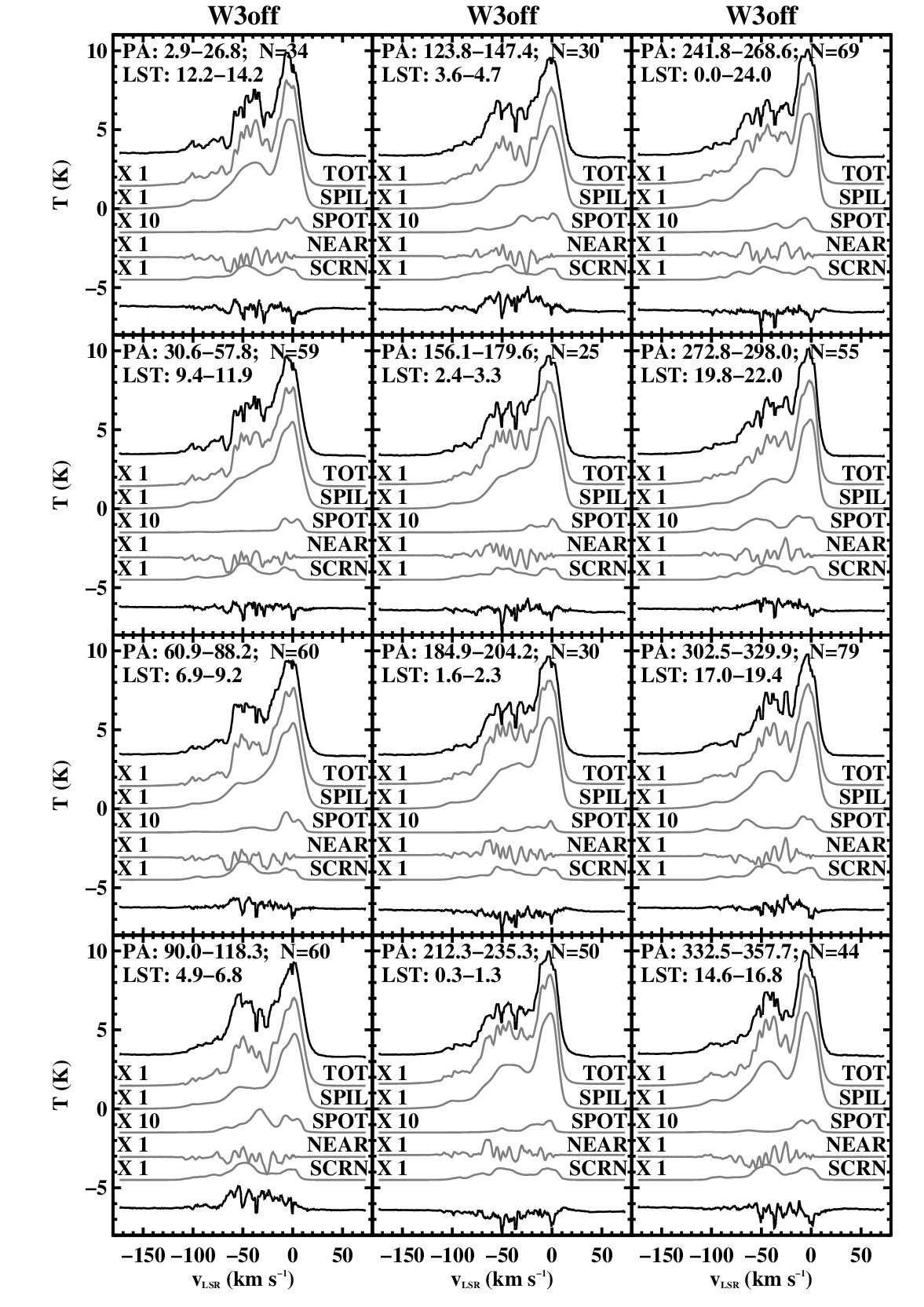}
  \end{center}
  \caption {Observed and predicted sidelobe contributions for W3off.
    Each panel shows the averages for $30^{\circ}$-wide bins of parallactic
    angle PA, with successive bins running down the page and then to the
    next column. The top curve (in black) in each panel is the total
    observed sidelobe contribution; the next curves, in gray, are the
    predicted sidelobe contributions; the bottom curve, in black, is the
    difference between the total observed and predicted contributions. For
    more details, see text (\S \ref{fitresults}).
    \label{w3offfit} }
\end{figure}

\begin{figure}[!p]
  \begin{center}
    \includegraphics[width=5in]{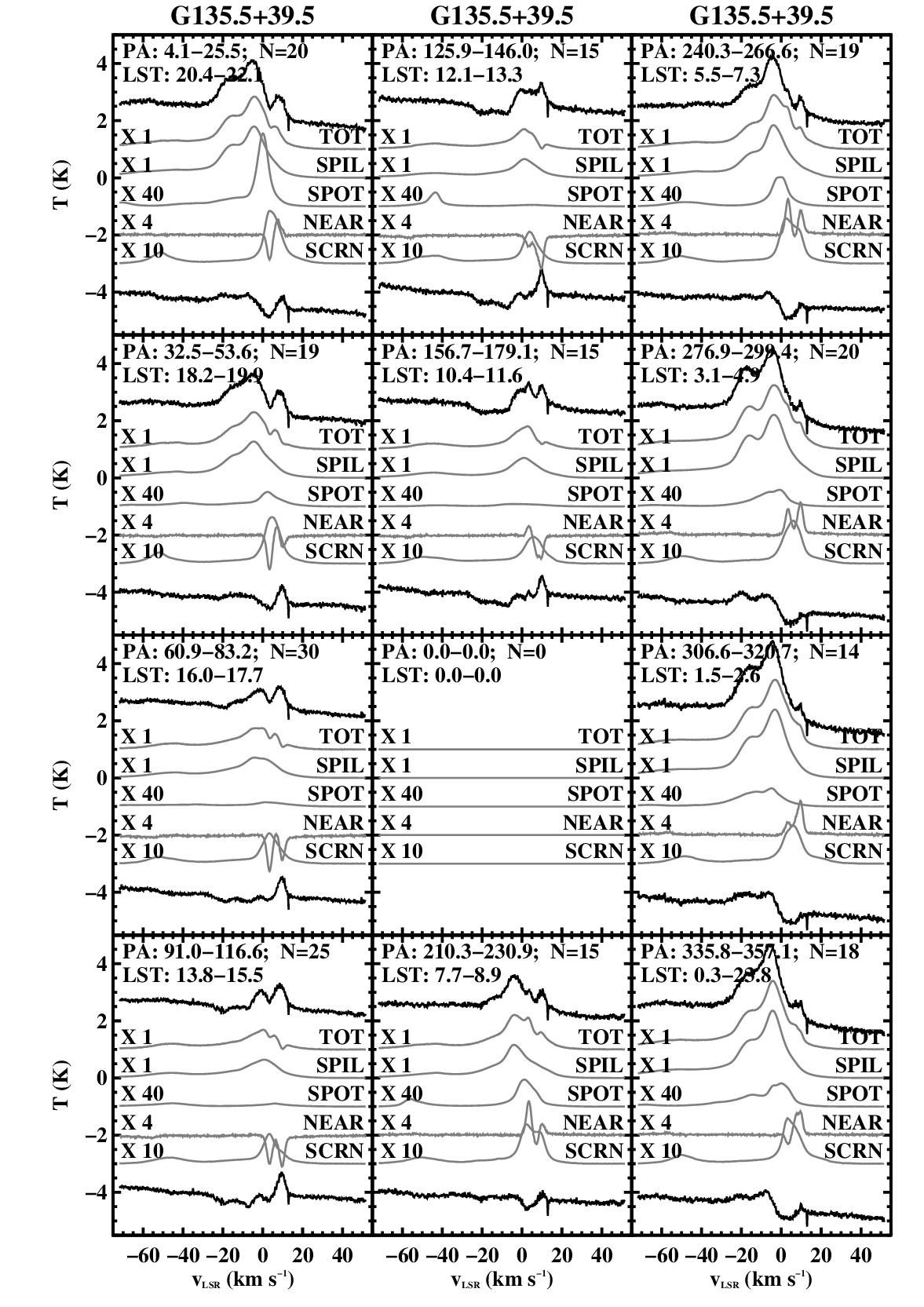}
  \end{center}
  \caption {Observed and predicted sidelobe contributions for G135.5$+$39.5.
    Each panel shows the averages for $30^{\circ}$-wide
    bins of parallactic angle PA, with successive bins running down the
    page and then to the next column. The top curve (in black) in each
    panel is the total observed sidelobe contribution; the next curves,
    in gray, are the predicted sidelobe contributions; the bottom
    curve, in black, is the difference between the total observed and
    predicted contributions. For more details, see text (\S
    \ref{fitresults}).
    \label{g135fit} }
\end{figure}

\begin{figure}[!p]
  \begin{center}
    \includegraphics[width=5in]{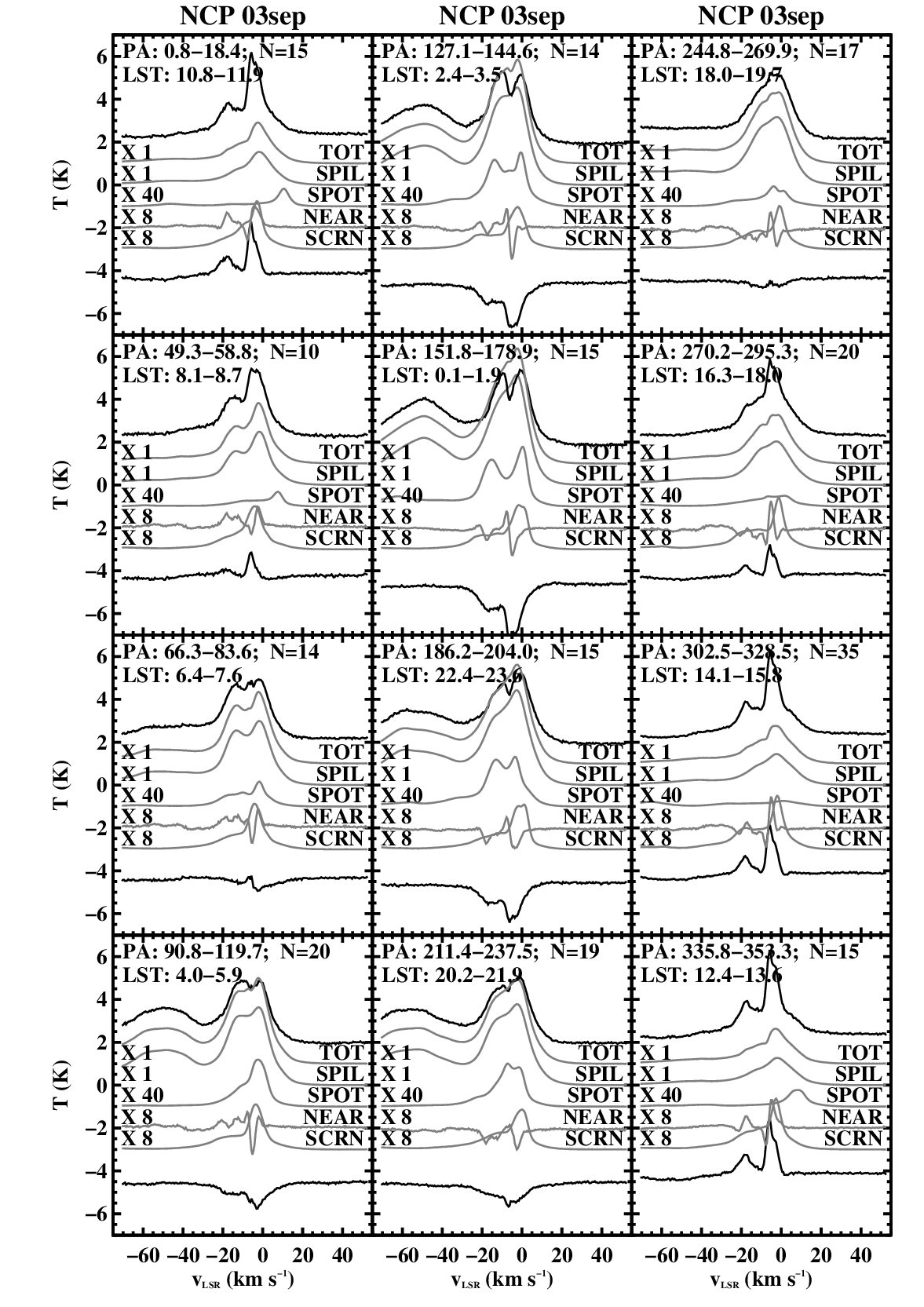}
  \end{center}
  \caption {Observed and predicted sidelobe contributions for the NCP data
    of 2003 September. Each panel shows the averages for $30^{\circ}$-wide
    bins of parallactic angle PA, with successive bins running down the
    page and then to the next column. The top curve (in black) in each
    panel is the total observed sidelobe contribution; the next curves,
    in gray, are the predicted sidelobe contributions; the bottom
    curve, in black, is the difference between the total observed and
    predicted contributions. For more details, see text (\S
    \ref{fitresults}).
    \label{ncp03sepfit} }
\end{figure}

Figures \ref{g139fit}--\ref{ncp03sepfit} show the application of the
above coefficients for the four sources. In each figure there are 12
panels, each showing averages for $30^{\circ}$-wide bins of parallactic
angle PA. The bins have nominal centers at (${\rm PA}=15^{\circ}$,
$45^{\circ}, ...$) running down the page and then to the next column;
annotations for each panel provide the actual ranges of PA and LST,
together with the number of spectra measured in each bin.  Within each
panel there are seven line profiles, which include data and residuals
(in black) and predicted components (in gray).  Some predicted profiles
are scaled up in intensity by the factors shown on the left; otherwise
they would not be discernible. Within each panel, from top to bottom the
profiles are: \begin{enumerate}

\item In black, the observed residual profile $T^{\rm resid}_{c,s} =
  T^{o}_{c,s}/G^{*}_{s} - T^{*}_{c}$, which is the total predicted
  contribution of GBT sidelobes. It is obtained by subtracting the
  derived ``true spectrum'' from each gain-corrected observed spectrum.

\item In gray, the sum $T^{\rm stray}_{c,s}$ of the four predicted
  coefficient contributions (``TOT''), which are the next four spectra.

\item In gray, the predicted spillover contribution (``SPIL'').

\item In gray, the predicted spot contribution (``SPOT'').

\item In gray, the predicted Nearin diffraction ring contribution (``NEAR''). This
  can be negative because we modelled the diffraction ring to have zero
  net gain; rather, it is a spatial difference.

\item In gray, the predicted screen contribution (``SCRN'').

\item In black, the difference between observed and predicted sidelobe
  spectra of (1) and (2) above, i.e., $\Delta T_{c,s} = T^{\rm resid}_{c,s}
  - T^{\rm stray}_{c,s}$. If the predicted spectra are accurate, then this
  difference is zero.
  
\end{enumerate}
         
For the two Galactic plane positions of Figures \ref{g139fit} and
\ref{w3offfit}, the difference $\Delta T_{c,s}$ is quite small. For
the high-latitude positions the difference has systematic nonzero
features. These amount to perhaps 1/4 of the observed residual profile
$T^{\rm resid}_{c,s}$. This reveals inadequacies in our modelling of the
sidelobes.
Examining these figures shows that the spillover contribution, which
covers large velocity ranges and is spectrally smooth, tends to
be well-predicted. Most of the residuals cover smaller velocity ranges
and have considerable spectral structure. 

For example, consider the results for G135.5$+$39.5. The spillover
contribution, which comes from large angular distances, covers wide
velocity ranges, from $\sim$$-30$ to $\sim$$+10$ km s$^{-1}$; in
contrast, most of the nonzero $\Delta T_{c,s}$ occurs in a smaller
velocity range typically centered on $\sim$$+5$ km s$^{-1}$. This narrower
range is where the line profile itself is strong. This implies that the
inadequacies lie close to the main beam, and in particular imply that
the situation is more complicated than the screen and diffraction-ring
models we use.

All of this strongly suggests that our predictions from the screen and
Nearin components, which lie close to the main beam and occupy small sky
areas, are not accurate. Obtaining this near-in beam structure requires
very sensitive and high-dynamic range maps of a strong, unresolved source.
The Sun is unsuitable because of its large angular diameter. The source Cas
A is suitable. However, while obtaining sensitivity requires integration
time---which is in principle possible---dealing with the required dynamic
range is very difficult. Classically, the dynamic-range problem has been
successfully solved only using an interferometric technique
\citep{hartsuijkerbdg72}.

\section{THE ACCURACY OF STOKES $V$ 21 CM LINE PROFILES}
\label{stokesv}

Above, in \S\ref{empiricali} and Figure \ref{ncppaper}, we found the
measured Stokes $V$ profiles at the NCP to change significantly with time.
This shows that the Stokes $I$ sidelobes are partially circularly
polarized.  Our main interest in using the GBT for 21 cm line work is
measuring magnetic fields using Zeeman splitting, which requires
measuring Stokes $V$.  The 21 cm line velocity changes with position.
Angular structure in the Stokes $V$ beam and sidelobes makes the two
circular polarizations sample slightly different velocities, and this
mimics Zeeman splitting.  In this section, we explore the seriousness of
this effect for the GBT.

\subsection{Stokes $V$ Sidelobes Related to the  Secondary Reflector}
\label{vsecondary}

\begin{figure}[!h]
  \begin{center}
    \includegraphics[width=6in]{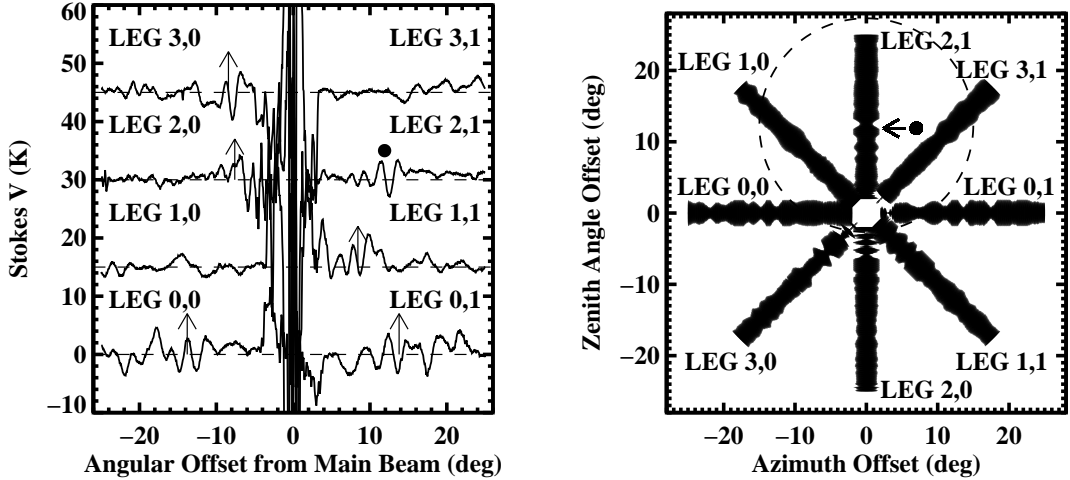}
  \end{center}
  \caption{Stokes $V$ analog for Figure \ref{sunscansi}. Four Stokes $V$ scans through
    the Sun in AZ/ZA great-circle offset coordinates.  In the left panel we
    plot the measured antenna temperatures versus the angular offsets from the
    main beam center. In the right panel, the azimuthal thicknesses of the
    lines are proportional to the measured temperatures; the dashed circle
    represents the approximate boundary of the secondary reflector as seen by
    the feed.. The begin and end points of the legs are labelled on each panel
    by ``,0'' and ``,1'', respectively.. On the left panel, the arrows lie
    $5^\circ$ outside the dashed circle. On both panels, the spot marks the
    Arago spot.
    \label{sunscansv}}
\end{figure}

Figure \ref{sunscansv}, left panel, exhibits Stokes $I$ versus AZ/ZA
great-circle offset coordinates for the same four scans through the Sun
that are shown in Figure \ref{sunscansi}. These data are for 2003 September 19.
The data for 2003 September 7 are similar, but not quite identical; we
are unsure whether the differences (which are not very big) are real.
Nevertheless, there exists real structure in the Stokes $V$ scans. We
mention three notable features: \begin{enumerate}

\item The spillover ring exhibits changes in Stokes $V$ on angular
scales ranging downward from about 2 degrees.  This structure is readily
visible as apparently random wiggles on the left panel of Figure
\ref{sunscansv}, particular for Leg 0.

\item The screen component---i.e., structure within a
  few degrees of the main beam---exhibits significant, rapidly varying
  Stokes $V$ structure. Especially prominent are Legs 2 and 0. The
  Stokes $V$ fractional polarizations are $\sim$25\% for the small
  negative ZA offsets portion of Leg 2 and range from $10\%$ to $20\%$
  for the other scans. Near beam center, Leg 0 exhibits a well-defined
  positive/negative lobe with peaks centered near offsets of a few
  degrees.
  
\item The Arago spot, which is centered near $\Delta {\rm ZA} = 12^\circ$
  on Leg 2 (marked with the black spot on Figures \ref{sunscansi} and
  \ref{sunscansv}), exhibits classical beam squint in Stokes $V$, with the
  Stokes $V$ intensity crossing zero where Stokes $I$ is maximum.
  Moreover, the spot's diffraction rings also exhibit beam squint. The
  fractional squint peak-to-peak ($I$,$V$) intensities are about $(67,7)$ K
  and $(10,2.5)$ K for the spot and its first sidelobe, respectively; these
  correspond to fractional polarizations ($\sim$10\%, 25\%). The spot's
  angular diameter is about 1 degree and the angular distance of the spot's
  sidelobe from its peak is about 2 degrees.

\end{enumerate}

\subsection{Stokes $V$ Structure of the Primary Beam} \label{vprimary}

\begin{figure}[!h]
  \begin{center}
    \includegraphics[width=6.0in]{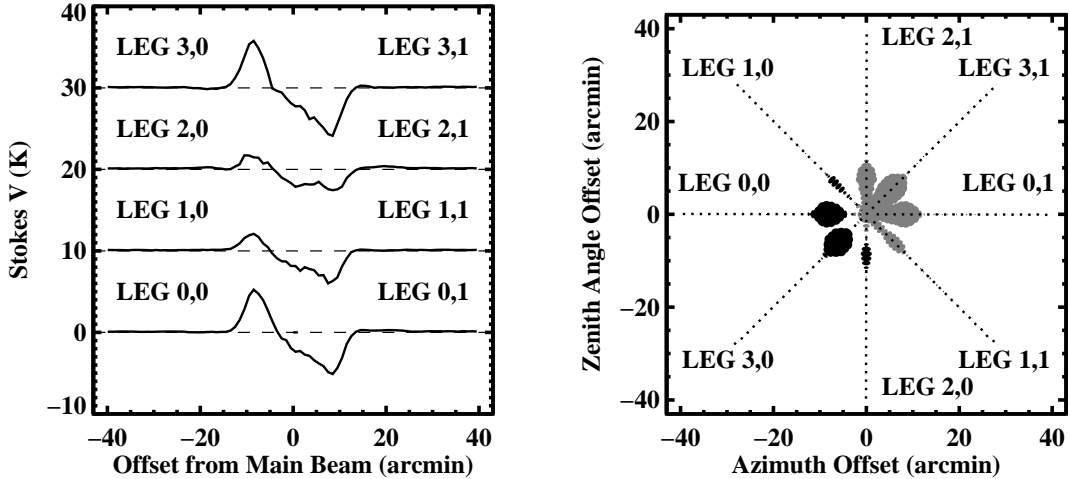}
  \end{center}
  \caption{Stokes $V$ analog of Figure \ref{casscansi}: Four Stokes $V$
    scans (``Legs'') through Cas A in AZ/ZA great-circle offset
    coordinates, showing the main beam squint. In the left panel we plot
    the measured antenna temperatures versus the angular offsets from the
    main beam center. In the right panel, the thicknesses of the lines are
    proportional to the measured temperatures; black is positive and gray
    is negative. The begin and end points of the legs are labelled on each
    panel.
    \label{casscansv}}
\end{figure}

Figure \ref{casscansv} is the Stokes $V$ analog of Figure
\ref{casscansi}. It represents the GBT Stokes $V$ main-beam diffraction
rings in two ways, with intensity versus position offset for four
observed legs in the left panel and a map in the right.  The main beam
squint is clear, with the center of the positive lobe centered at about
$\phi \approx 205^\circ$. The data were taken at ${\rm ZA} \approx
26^\circ$; the data of Figure \ref{getsquint} imply $\phi \approx
250^\circ$ at ${\rm ZA} \approx 26^\circ$. This difference in $\phi$,
about $45^\circ$, is uncomfortably large and suggests that the squint
direction depends on more than just ZA.

The first diffraction ring, which is centered $\sim$$20^\circ$ from beam
center, exhibits a weak Stokes $V$ signature which is roughly opposite in
sign to (and is more complicated than) the main beam's squint pattern. We
don't show it because the signal/noise is too poor to derive the details.

\subsection{Stokes $V$ Problems for Each Source}

\begin{figure}[!p]
  \begin{center}
    \includegraphics[width=5in]{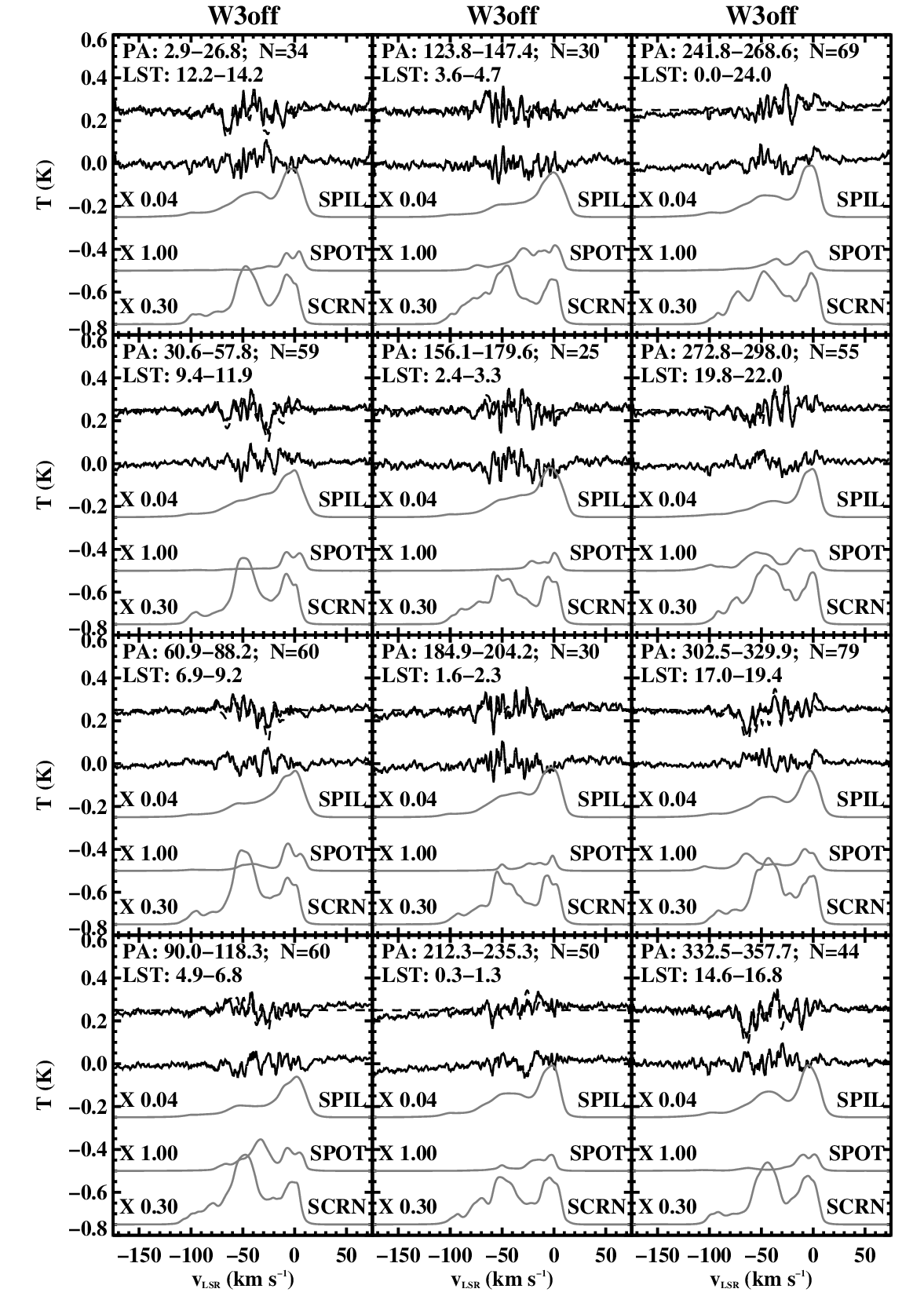}
  \end{center}
  \caption {Stokes $V$ (top two profiles) and predicted Stokes $I$ sidelobe
    contributions for W3off.  Each panel shows the averages for
    $30^{\circ}$-wide bins of parallactic angle PA, with successive
    bins running down the page and then to the next column. The top
    curve in each panel is the bin-average Stokes $V$ profile; the next
    is that profile, corrected for main-beam squint. The next three
    profiles, in gray, are the predicted Stokes $I$ contributions from
    spillover (``SPIL''), the Arago spot (``SPOT''), and the screen
    (``SCRN''), respectively, multiplied by the factors given on the
    left-hand side for each profiles.
    \label{w30ffv}}
\end{figure}

\begin{figure}[!p]
  \begin{center}
    \includegraphics[width=5in]{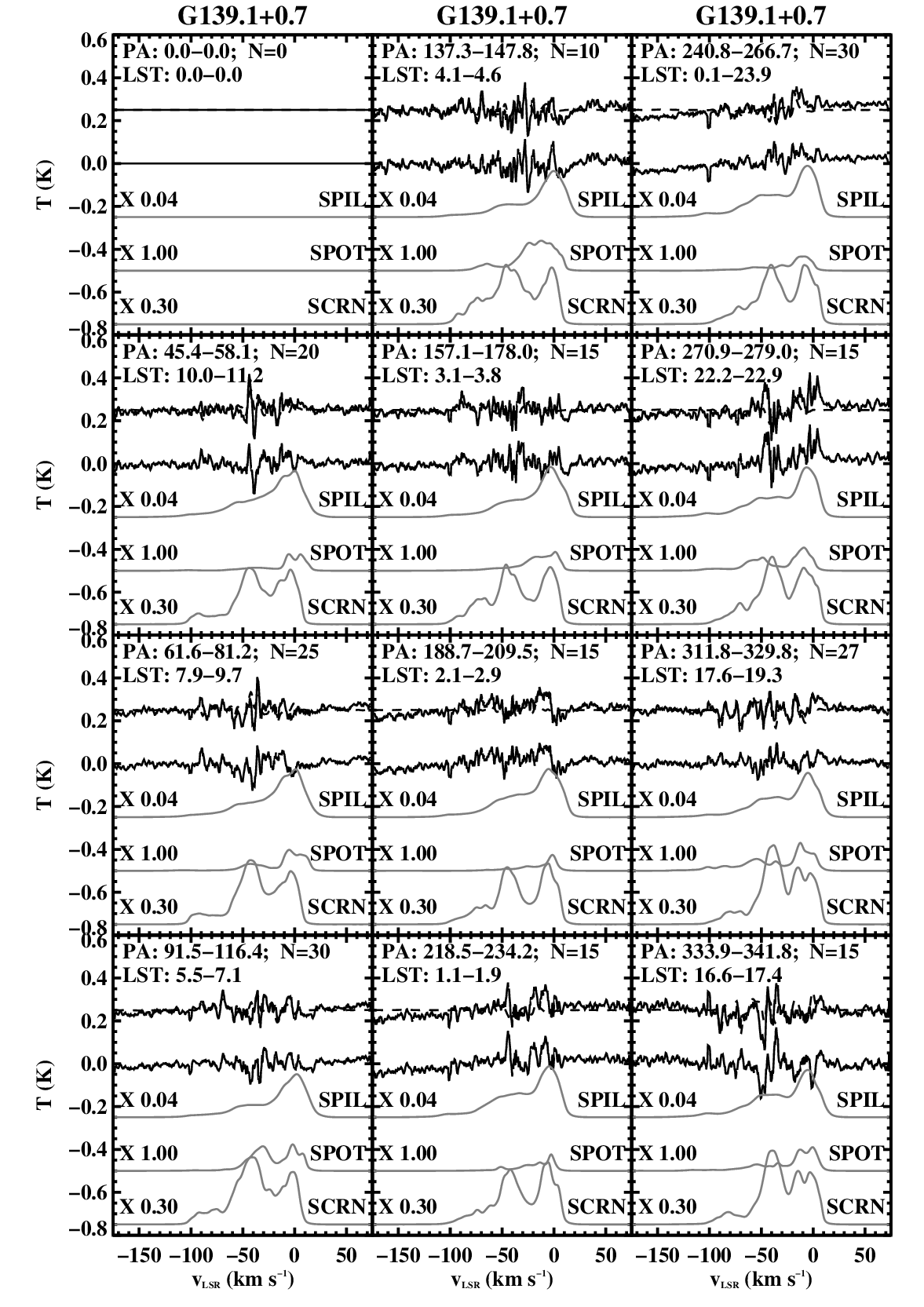}
  \end{center}
  \caption {Stokes $V$ (top two profiles) and predicted Stokes $I$ sidelobe
    contributions for G139.1$+$0.7.  Each panel shows the averages for
    $30^{\circ}$-wide bins of parallactic angle PA, with successive
    bins running down the page and then to the next column. The top
    curve in each panel is the bin-average Stokes $V$ profile; the next
    is that profile, corrected for main-beam squint. The next three
    profiles, in gray, are the predicted Stokes $I$ contributions from
    spillover (``SPIL''), the Arago spot (``SPOT''), and the screen
    (``SCRN''), respectively, multiplied by the factors given on the
    left-hand side for each profiles.
    \label{g139v}}
\end{figure}

\begin{figure}[!p]
  \begin{center}
    \includegraphics[width=5in]{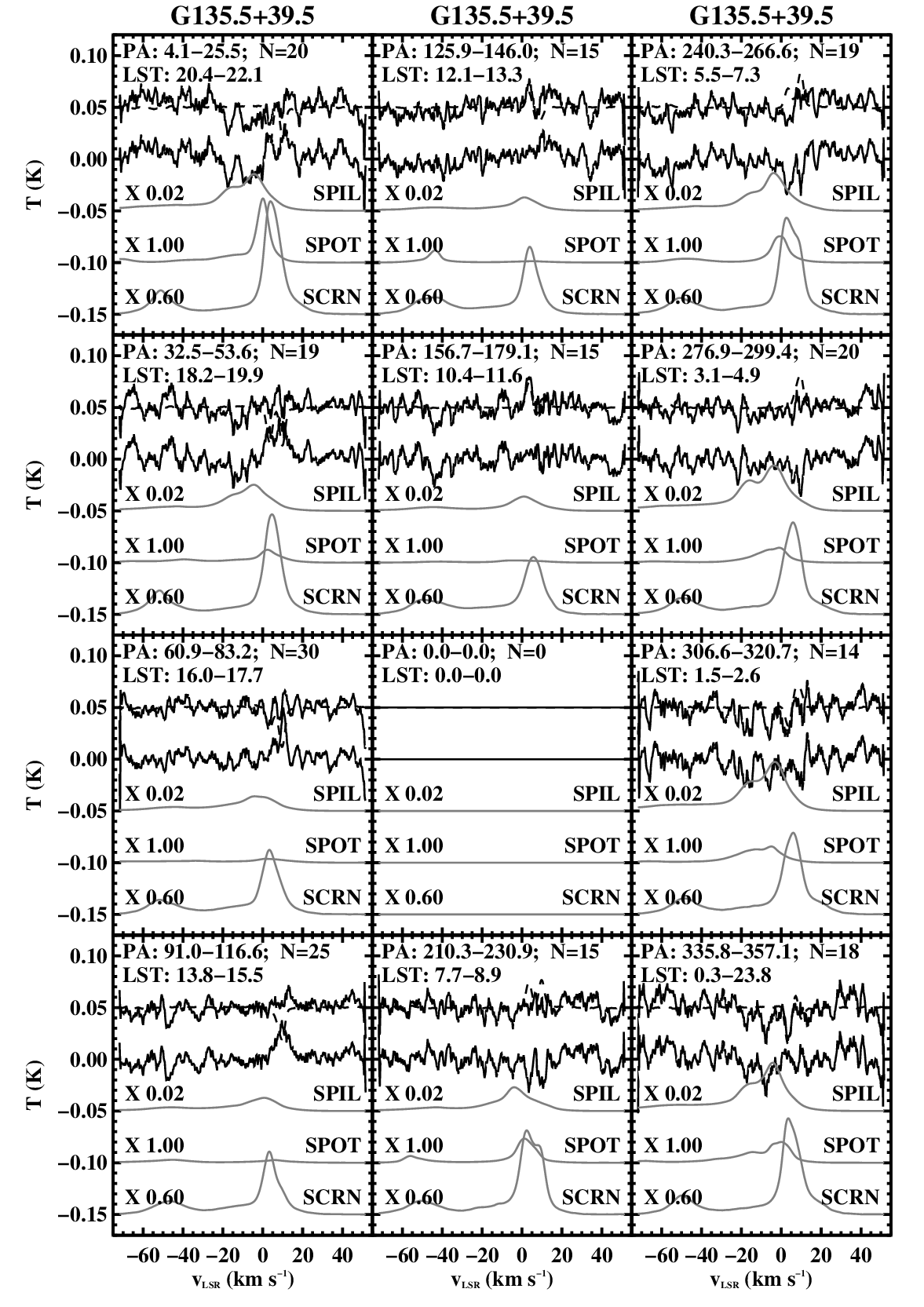}
  \end{center}
  \caption {Stokes $V$ (top two profiles) and predicted Stokes $I$ sidelobe
    contributions for G135.5$+$39.5.  Each panel shows the averages for
    $30^{\circ}$-wide bins of parallactic angle PA, with successive
    bins running down the page and then to the next column. The top
    curve in each panel is the bin-average Stokes $V$ profile; the next
    is that profile, corrected for main-beam squint. The next three
    profiles, in gray, are the predicted Stokes $I$ contributions from
    spillover (``SPIL''), the Arago spot (``SPOT''), and the screen
    (``SCRN''), respectively, multiplied by the factors given on the
    left-hand side for each profiles.
    \label{g135v}}
\end{figure}

\begin{figure}[!p]
  \begin{center}
    \includegraphics[width=5in]{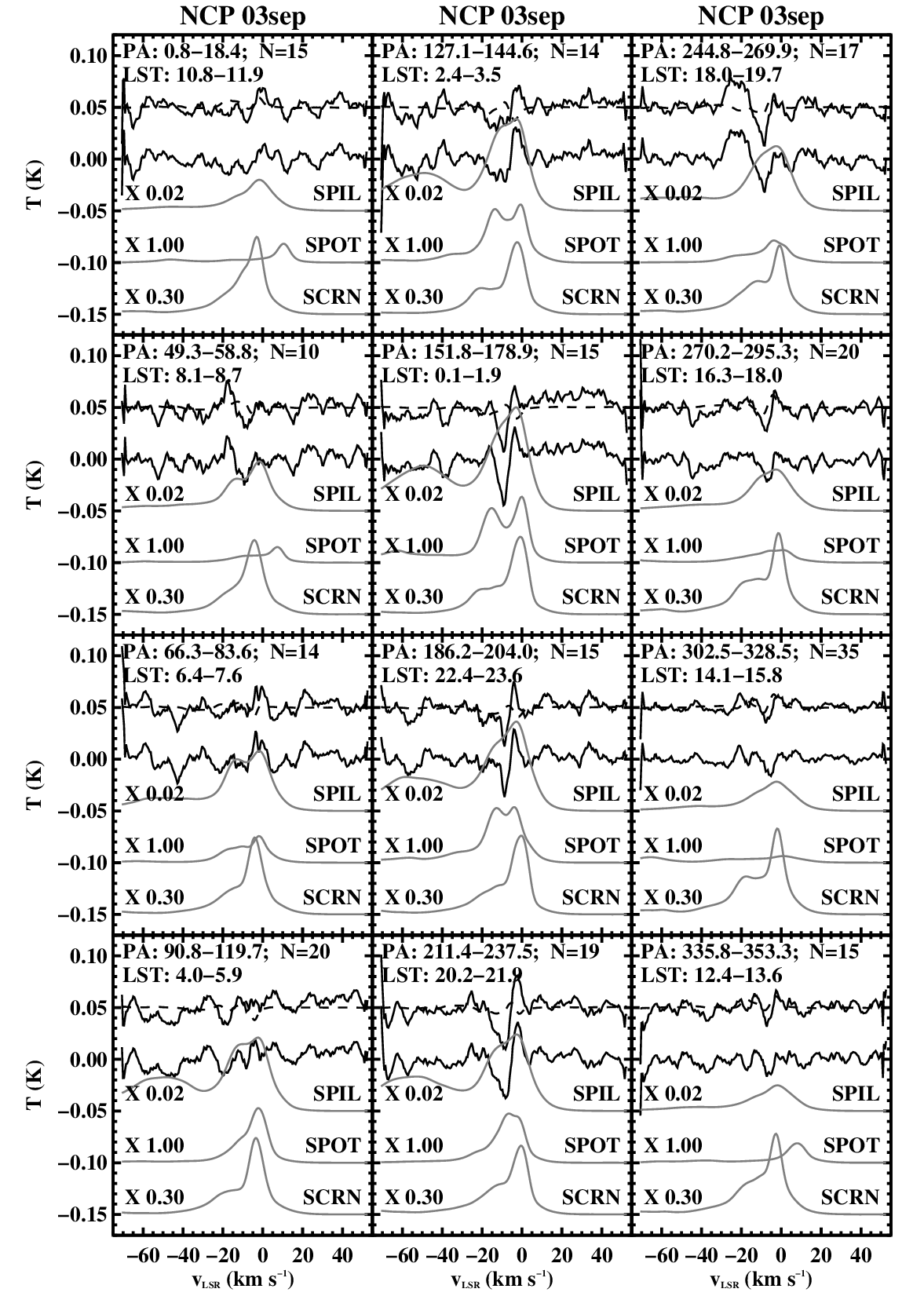}
  \end{center}
  \caption {Stokes $V$ (top two profiles) and predicted Stokes $I$ sidelobe
    contributions for the NCP.  Each panel shows the averages for
    $30^{\circ}$-wide bins of parallactic angle PA, with successive
    bins running down the page and then to the next column. The top
    curve in each panel is the bin-average Stokes $V$ profile; the next
    is that profile, corrected for main-beam squint. The next three
    profiles, in gray, are the predicted Stokes $I$ contributions from
    spillover (``SPIL''), the Arago spot (``SPOT''), and the screen
    (``SCRN''), respectively, multiplied by the factors given on the
    left-hand side for each profiles.
    \label{ncpv}}
\end{figure}

Figures \ref{w30ffv}--\ref{ncpv} present PA-binned Stokes $V$ profiles
together with the predicted Stokes $I$ profiles from the four sidelobe
contributions for the four sources.  In each figure there are 12 panels,
each showing averages for $30^{\circ}$-wide bins of parallactic angle PA.
The bins have nominal centers at (${\rm PA}=15^{\circ}$, $45^{\circ}, ...$)
running down the page and then to the next column; annotations for each
panel provide the actual ranges of PA and LST, together with the number of
spectra measured in each bin.

Within each panel there are five line profiles, which include two Stokes
$V$ profiles (in black) and three predicted Stokes $I$ profiles for
different sidelobe components (in gray).  The predicted profiles are scaled
in intensity by the factors shown on the left. The spillover component is
generally scaled down by a factor $\sim$20--40 and the screen component by
a factor $\sim$2--3; otherwise they would not fit on the panels. Within
each panel, from top to bottom the profiles are:
\begin{enumerate}

\item In black, the Stokes $V$ profile.

\item In black, the Stokes $V$ profile corrected for main-beam squint.

\item In gray, the predicted Stokes $I$ spillover contribution (``SPIL''),
  scaled down by a factor $\sim$20--40 as indicated.

\item In gray, the predicted Stokes $I$ spot contribution (``SPOT'').

\item In gray, the predicted Stokes $I$ screen contribution (``SCRN''),
  scaled down by a factor $\sim$2--3 as indicated.

\item We don't show the predicted Nearin contribution because it covers
  roughly the same velocity range as the screen contribution, but has
  more complicated structure. 

\end{enumerate}
         
First, consider the main-beam squint correction. In some bins it makes a
huge improvement. Examples include the ${\rm PA}=272.8^\circ {\rm -}
298.0^\circ$ bin for W3off, the ${\rm PA}=311.8^\circ {\rm -} 329.8^\circ$ bin
for G139.1$+$0.7, and the ${\rm PA}=125.9^\circ {\rm -} 146.0^\circ$ bin for
G135.5$+$39.5. In most cases, though, its correction represents only a
small portion of the time-variable systematic errors in the Stokes $V$
profiles.

We conclude that the circularly-polarized portions of the four spillover
contributions are responsible for most of the Stokes $V$
inaccuracies. Figures \ref{w30ffv}--\ref{ncpv} provide some clues
about which components are serious contributors to these errors. We
consider, first, the Spillover and Spot contributions. \begin{enumerate}

\item {\it The Spillover component}. Generally, the Stokes $I$ Spillover
  profile covers a large velocity range. By inspection, this allows us to
  see that the Stokes $V$ problems do not scale very well with the Stokes
  $I$ spillover intensity. For example, the ${\rm PA}=123.8^\circ {\rm -}
  147.4^\circ$ bin for W3off has severe Stokes $V$ wiggles for $v_{\rm
    \scriptstyle LSR} \lesssim -25$ km s$^{-1}$, where the Stokes $I$
  Spillover profile is weak; more generally for this source, the Stokes $I$
  Spillover profile changes quite a bit in intensity for $v_{\rm
    \scriptstyle LSR} \lesssim -25$ km s$^{-1}$, but the wiggle amplitudes
  in the Stokes $V$ spectra don't track these changes. The same general
  statement applies to G139.1$+$0.7.  Similarly, for the two high-latitude
  sources the Stokes $I$ spillover amplitude varies considerably with PA,
  but the Stokes $V$ wiggles don't track these variations.

\item {\it The Spot component}. Generally, the Stokes $I$ Spot profile
  covers a smaller velocity range than the Spillover profile and is less
  spectrally smooth. This makes it easier to see that the Stokes $V$
  problems are not well-correlated with the Stokes $I$ spot profile
  intensity. Clear examples include the ${\rm PA}=156.1^\circ {\rm -}
  179.6^\circ$ bin for W3off, the ${\rm PA}=157.1^\circ {\rm -}
  178.0^\circ$ bin for G139.1$+$0.7, and many of the bins for the two
  high-latitude sources.

\item  We conclude that most of the Stokes $V$ wiggles do not come from the
either the Spillover or the Spot lobe.

\end{enumerate}

By the process of elimination, this leaves the Screen and Nearin
components. Consider, first, the possible contribution to Stokes $V$ from
the Nearin component, i.e., the diffraction rings. We concluded above in
\S \ref{diffraction} that the integrated response of the diffraction
rings is $\lesssim 1\%$ that of the main beam. In \S \ref{vprimary} we
found that the diffraction rings are almost unmeasurable in Stokes $V$ and
are $\lesssim 20$ times weaker than the main-beam squint. We have seen
that the main-beam squint accounts for only a fraction of the observed
Stokes $V$ wiggles. The contribution from the diffraction rings should be
smaller. We conclude that the diffraction rings contribute
insignificantly to the Stokes $V$ wiggles.

\subsection{The Culprit Is the Screen Component}

Again by process of elimination, we are left with the Screen component as
the culprit that produces the extraneous Stokes $V$ wiggles. This
conclusion makes quantitative sense. First, the integrated beam response of
the Screen component is roughly $1\%$ the main beam's (\S \ref{details};
Table \ref{lscoeffs}). This seems small, but the Stokes $V$ wiggles are of
order $10^{-4}$ the Stokes $I$ amplitude, and the Screen component is much
more highly polarized than the main beam. Above, in \S \ref{vsecondary}, we
concluded its Stokes $V$ fractional polarization reaches $\sim$25\%.
Moreover, this polarization changes over the size of the Screen component,
which is a ring of radius $\sim$$2.8^\circ$ (eq.\ [\ref{screeneqn}]). Thus,
it samples {\it changes} in the 21 cm profiles over much larger angles than
the main beam does.

Unfortunately, our estimate of the screen beam is very rough. It is very
difficult to accurately measure the screen component---not only in
Stokes $I$, as we remarked above in \S \ref{fitresults}, but even more
so in Stokes $V$. This means that we cannot present any hard and fast
rule about its effect. We can only offer a rule of thumb, which we base
on comparing the Stokes $V$ wiggle amplitude with the predicted Stokes
$I$ screen profile of Figures \ref{w30ffv}--\ref{ncpv}. In these
figures, the Stokes $V$ wiggle is of order 0.2 times the predicted
screen Stokes $I$ profile. So we can offer the following:
\begin{enumerate}

\item For Stokes $V$ measurements in narrow ranges of parallactic angle
  PA, results can be trusted if they are stronger than $\sim$0.2 the
  predicted Stokes $I$ screen component.

\item Averaging over larger ranges of parallactic angle PA helps
  greatly. If one has sampling of the full range of PA, then the
  Stokes $V$ results are much more reliable.
\end{enumerate}


\section{SUMMARY AND PERSPECTIVE}
\label{conclusion}

\subsection{Summary}

We sampled the GBT sidelobes and used structural and physical
considerations to define four sidelobe components. Three of these are
related to the feed's illumination of the secondary reflector: the
spillover, the spot, and the screen components (\S \ref{sidelobeobsi}). A
fourth component, which we call the nearin component, is related to the
diffraction rings (\S \ref{diffraction}). We applied a least-squares
procedure (\S \ref{lsfits}, \S \ref{lsresults}) to full hour-angle coverage
of 21 cm line data for four positions to obtain the amplitude of these
components, and also to obtain the errors in calibrated gains for each
spectrum and the true profile for each position.

We find the following items of importance for accurate 21 cm line
profiles measured by the GBT: \begin{enumerate}

\item We find intrinsic differences between calibrated intensity for GBT
  versus LAB profiles at the $\sim$60\% level (Table \ref{intratio}),
  which we attribute to the higher main-beam efficiency of the GBT; this,
  in turn, is a direct result of its clear, unblocked aperture.

\item We find systematic time variations i n the calibrated gain (Figure
  \ref{allgains}), which means that GBT 21 cm line profile intensities
  cannot be relied on at the $10\%$ level.

\item The combination of the spillover, spot, and screen sidelobes has an
  area-integrated beam response which is almost $10\%$ that of the main
  beam (Table \ref{lscoeffs}). These can produce significant inaccuracy in
  observed Stokes $I$ 21 cm line profiles (stray radiation).

\item We use the solutions for the sidelobe beam responses, together
  with the LAB survey, to predict and remove the 21 cm
  line contaminants with fair accuracy (\S \ref{fitresults}).

\item The distant sidelobes also contribute to Stokes $V$ 21 cm line
  profiles (\S \ref{stokesv}). In particular, the screen component
  contributes a currently uncorrectable contaminant to the Stokes $V$
  profiles. This has amplitude $\sim$20\% of its Stokes $I$ counterpart, so
  GBT Stokes $V$ 21 cm line profiles cannot be trusted at this level unless
  they are averaged over significant ranges of parallactic angle.

\end{enumerate}

In summary, 21 cm line profiles in both Stokes $I$ and Stokes $V$ contain
significant contaminants. For Stokes $I$, these can be removed with some
success with the procedure presented in this paper. Uncertainties in the
predicted contaminants are dominated by our inaccurate knowledge of the screen
component. This can, in principle, be measured more accurately, but doing
so requires high dynamic range measurements of an unresolved source; these
are probably feasible only using the interferometric technique of
\citet{hartsuijkerbdg72}.

\subsection{Perspective}

We find it ironic that, historically, one of the arguments for the
clear-aperture design of the GBT was the enhanced accuracy of 21 cm line
profiles. The purity of its clear-aperture design is spoiled by the use of
a Cassegrain system, with its secondary reflector. Upon reflection, we see
that using the GBT in a prime-focus mode would retrieve most, perhaps all,
of this purity. One would pay for this with a large beam squint in Stokes
$V$---but this would be a predictably reliable and straightforward
correction. It seems to us that using the GBT in prime-focus mode would
provide more accurate 21 cm line profiles in both Stokes $I$ and $V$ than
performing a series of uncertain and perhaps time-variable corrections to
profiles measured with the Cassegrain system.


\acknowledgements This work was supported in part by the NSF grant
AST 0406987, as well as through awards GSSP 02-0011, 05-0001, 05-0003,
05-0004, and 06-0003 from the NRAO.  We greatly appreciate the help of NRAO
staff members Roger Norrod, Rick Fisher, Karen O'Neil, and Carl Bignell.
We also thank Peter Kalberla for enlightening discussions.

{\it Facilities:} \facility{GBT}


\clearpage

\bibliographystyle{apj}

\newcommand{\noopsort}[1]{}

\end{document}